\documentclass[sn-mathphys,Numbered]{sn-jnl}

\usepackage{graphicx}%
\usepackage{multirow}%
\usepackage{amsmath,amssymb,amsfonts}%
\usepackage{amsthm}%
\usepackage{mathrsfs}%
\usepackage[title]{appendix}%
\usepackage{xcolor}%
\usepackage{textcomp}%
\usepackage{manyfoot}%
\usepackage{booktabs}%
\usepackage{algorithm}%
\usepackage{algorithmicx}%
\usepackage{algpseudocode}%
\usepackage{listings}%

\begin{document}

\title{HNL see-saw: lower mixing limit and pseudodegenerate state}

\author*{\fnm{Igor} \sur{Krasnov}}\email{iv.krasnov@physics.msu.ru}

\affil*[1]{\orgname{Institute for Nuclear Research of Russian Academy of Sciences}, \orgaddress{\street{prospekt 60-letiya Oktyabrya 7a}, \city{Moscow}, \postcode{117312}, \country{Russia}}}

\abstract{Heavy Neutral Leptons are popular hypothetical particles, first introduced as a way to explain neutrino oscillations, and since then extensively studied in relation to many other aspects of physics beyond the Standard Model. They also serve as viable targets for direct experimental searches, being effectively described only by HNL mass and mixing with each neutrino flavor. Motivated by this, we study the lower theoretical boundary for mixing with a specified flavor in two and three HNLs cases. We find the connection of this limit with the effective neutrino mass appearing in neutrinoless double beta decay (and similar expressions for mixing with muon and tau neutrino). In two HNLs case, there is a rather strict relation between mixing of different HNLs with the same neutrino flavor. We find that existing exclusion regions and their expected expansions in the near future are all described by a certain limit. We call that limit pseudodegenerate state and find its relation to the symmetrical limit, already studied in the literature. We also study pseudodegenerate state and conditions under which it is achieved in three HNLs case.}

\keywords{beyond standard model, neutrino physics, high energy physics}
\maketitle
\flushbottom

\section{Introduction}
Heavy Neutral Leptons (HNLs) are one of the primary candidates for the extension of the Standard Model (SM), originally suggested as a solution to the neutrino mass scale problem in a so-called seesaw mechanism\,\cite{Minkowski:1977sc,Gell-Mann:1979vob,Mohapatra:1979ia,Yanagida:1980xy,Schechter:1980gr} and since then found to be capable of contributing to the baryon asymmetry of the Universe\,\cite{Fukugita:1986hr,Harvey:1990qw,Luty:1992un,Plumacher:1996kc,Covi:1996wh,Flanz:1996fb,Canetti:2012zc} or playing the role of dark matter\,\cite{Dodelson:1993je,Gorbunov:2014ypa}.
HNLs are unchanged under SM gauge groups, serving as right-handed Majorana counterparts for neutrinos, which earns them their other name, \emph{sterile neutrinos}.
We usually only consider scenarios where all sterile neutrinos have masses~$\gg$~eV, and, therefore, prefer the usage of the term ``heavy neutral leptons''.
The results obtained in this paper can be applied to sterile neutrinos with masses~$\sim$~eV that can explain reactor and gallium anomalies that continue to attract the interest of researchers~\cite{Barinov:2021mjj,Serebrov:2023vfo}, but a comprehensive analysis of that area falls outside the scope of this work.
Additionally, in this paper we limit ourselves to seesaw type-I mechanism~\cite{Minkowski:1977sc,Gell-Mann:1979vob,Mohapatra:1979ia,Yanagida:1980xy,Schechter:1980gr}.
Its Lagrangian can be written as:
\begin{align}
\mathcal{L}=&\mathcal{L}_{SM} + \mathnormal{i} \bar{N}_I\gamma^\mu \partial_\mu N_I - \Big(\frac{1}{2} M_I \bar{N}^c_I N_I + \hat{Y}_{\alpha I}\bar{L}_\alpha \tilde{H} N_I +h.c.\Big), \label{eq:lagrangian}
\end{align}

From the phenomenological point of view, HNLs can be fully described by its mass $M_I$ and mixing to the neutrino sector $|U_{\alpha I}|$, where $U = \frac{\mathnormal{v}}{\sqrt{2}}M_I^{-1} Y, \alpha \in \{ e, \mu, \tau\}$.
The original seesaw mechanism had an extremely large scale of HNL mass, close to Grand Unification scale $M_I\sim 10^{15}$ GeV, but it was shown that HNLs with masses $M_I \lesssim 1$ GeV but feeble mixing $|U_{\alpha i}| \ll 1$ can just as effectively serve that role~\cite{Abdullahi:2022jlv}.
Mixing with the neutrino sector implies that, if kinematically allowed, HNLs can be produced in weak interactions and later decay into SM particles.
Therefore many collider and beam-dump experiments, such as E949\cite{E949:2014gsn}, NA62\cite{NA62:2020mcv}, T2K\cite{T2K:2019jwa}, TRIUMF\cite{Britton:1992xv}, PIENU\cite{PIENU:2017wbj} and others place limits on HNL mixing.

Traditionally, the minimal seesaw mixing that is consistent with up-to-date active neutrino sector experimental data is taken to be a line $U^2 \geq \sum_{\alpha j} |U_{\alpha i}|^2 \geq \frac{m_\nu}{M_{max}}$, where active neutrino mass is approximated as $m_\nu \approx \sqrt{\Delta m_{atm}^2}$~\cite{Abdullahi:2022jlv}.
Another approach is to consider specific scenarios where only certain elements of the mixing matrix are designated as important~\cite{Gorbunov:2013dta,Krasnov:2018odt}.
The goal of this article is to study the strict analytical minimal value for $|U_{\alpha i}|^2$, different for each active neutrino flavour, and how it relates to existing and upcoming experimental bounds.
We also find and study a specific limit, that, as we find, is usually realised near these experimental bounds, which we call the pseudodegenerate HNLs state.

We start with a description of the general approach to HNL physics with our study for the two HNLs case in section~\ref{sec:2HNL} and then move on to a more general three HNLs case in section~\ref{sec:3HNL}.
We summarize our findings in the conclusion~\ref{sec:conclusions}.
In the first part of the appendix~\ref{sec:appendix} we provide some additional information on  active neutrino mixing parameters and corresponding experimental data we use throughout the paper, as well as provide analytic calculations for heavy neutral lepton sector in the second part~\ref{app:HNL}.

\section{Two HNLs case}
\label{sec:2HNL}

For our purposes it is convenient to adopt Casas-Ibarra parametrization \cite{Casas:2001sr} of HNL mixing angle:
\begin{align}
\label{eq:U}
&U= \frac{\mathnormal{v}}{\sqrt{2}}M_I^{-1} Y = \mathnormal{i}M_I^{-\frac{1}{2}} R m_\nu^{\frac{1}{2}} \times diag\{e^{i \frac{\alpha_1}{2}}; e^{i \frac{\alpha_2}{2}}; 1\}\,U_{PMNS}^\dagger.
\end{align}
Here $m_\nu \equiv diag\{m_1, m_2, m_3 \}$ and $M_I \equiv diag\{M_1, M_2, ... \}$. $U_{PMNS}$ is Pontecorvo-Maki-Nakagawa-Sakata matrix and $\alpha_1, \alpha_2$ are active neutrino Majorana phases. Matrix $R$ is a complex orthogonal matrix, $R^T R =1$.

Active neutrinos gain their mass after the electroweak symmetry breaking via the seesaw mechanism \eqref{eq:lagrangian}: $m_\nu = U M_I U^T$.
Because the HNL mass matrix has $N$ eigenstates, mixing with HNLs, in that case, provides $N$ eigenvalues to the active neutrino mass matrix.
Observed neutrino oscillation phenomena dictate that at least two active neutrinos have different nonzero masses, but don't fix neutrino mass scale~\cite{ParticleDataGroup:2024cfk}.
Therefore, the minimal number of HNLs that can explain all existing neutrino observations is two, which inevitably makes the lightest active neutrino massless.
The case of three HNLs is less restrictive on the active sector and is a bit more favored from the theoretical standpoint: for instance, many grand unification theories (such as $SO(10)$ models, where HNLs are also usually responsible for baryon asymmetry of the Universe~
\cite{Fukugita:1986hr,Harvey:1990qw,Luty:1992un,Plumacher:1996kc,Covi:1996wh,Flanz:1996fb}) include the same number of right-handed neutrinos as they have left-handed neutrinos.
It is possible to introduce four or more HNLs into theory, but due to the overabundance of free parameters it is rarely done without the introduction of specific symmetries of some kind, and these cases fall outside the scope of this work.

Two HNLs case was extensively studied in literature~\cite{Abdullahi:2022jlv}, and in this work we only focus our attention on two aspects: the minimal HNL mixing with specified neutrino flavor and a specific case we call pseudodegenerate state.

\subsection{Lower limit for mixing with specific flavor}
There is enough freedom in sterile sector to make any element of mixing matrix equal to zero, meaning that any one HNL can be set to not mix with a chosen active neutrino flavour.
The reduction in degrees of freedom greatly restricts the possible values of mixing of the other HNL with the same neutrino flavour.
%But to simultaneously achieve via see-saw type I mechanism the observed values of neutrino mixing angles and squared mass differences, one inevitably greatly restricts the possible value of the other HNL mixing with the same neutrino flavour.
%It is convenient to study contributions of HNLs towards mixing with one chosen neutrino flavour $\alpha$, for example $e$.
%One can choose parameters of HNL sector in such a way so that one HNL doesn't mix with said flavor.
We can omit peculiarities of distribution of mixing between different HNLs by studying their summary contribution to the mixing with chosen neutrino flavour $U_{sum, \alpha}^2 = |U_{1\alpha}|^2 + |U_{2\alpha}|^2$.
%The sum of contributions of both HNL to the mixing with neutrino flavour $U_{sum, \alpha}^2 = \frac{1}{M_1} |\Gamma_{11}|^2 + \frac{1}{M_2} |\Gamma_{21}|^2$ becomes the crucial parameter for calculating the impact of HNLs on mixing with active neutrino.
This expression has direct experimental meaning, for example in searches of "missing signal" that's contributed to HNL substituting active neutrinos in some dacays and scatterings (where it is kinematically allowed).
We analyse the expression $U_{min, \alpha}^2 = \left(U^2_{sum, \alpha}\right)_{min}$ minimized over the sterile sector parameters.
Obviously, \emph{at least one} HNL mixing has a greater or equal to value to the half-sum: $max\{|U_{1\alpha}|^2, |U_{2\alpha}|^2\} \geq \frac{1}{2} U^2_{min, \alpha}$.
%To get the minimal value above which must lay , we study the behaviour of the half-sum of mixing of two HNL (we start with the example of mixing with $\nu_e$): $U^2_e = \frac{1}{2 M_1} |\Gamma_{11}|^2 + \frac{1}{2 M_2} |\Gamma_{21}|^2$.

That expression can be analytically found (see appendix \ref{app:2HNL}):
\begin{align}
\label{eq:2HNL_min}
U^2_{min, \alpha} =&  \frac{1}{M_{max}} |m_1 U^{\dagger^2}_{{PMNS}_{1\alpha}} e^{i\alpha_1} + m_2 U^{\dagger^2}_{{PMNS}_{2\alpha}} e^{i\alpha_2}+ m_3 U^{\dagger^2}_{{PMNS}_{3\alpha}}| \equiv \frac{|m_{\alpha\alpha}|}{M_{max}}
\end{align}
These results hold true for both normal hierarchy ($m_1=0$) and inverted hierarchy ($m_3=0$) cases. 
If HNLs doesn't have a degenerate mass term ($M_1 \neq M_2$), minimal mixing with a specific active neutrino flavour is achieved when only one the heavier of the two HNLs is mixing with it. When $M_1= M_2$ the expression $U_{sum, \alpha}^2 = |U_{1\alpha}|^2 + |U_{2\alpha}|^2$ has the same minimal value, but both matrix elements can have nonzero value.

In case of mixing with electron flavour, the expression appearing in \eqref{eq:2HNL_min} is a known entity that appears as an effective neutrino mass $m_{ee}$ in neutrinoless double beta decay searches \cite{ParticleDataGroup:2024cfk}.
Expressions $m_{\mu\mu}$ and $m_{\tau\tau}$, much like $m_{ee}$, can appear in rare lepton number violating decays, such as $K^- \to \pi^+ \mu^- \mu^-$ or $B^- \to K^+ \tau^- \tau^-$ and other $\Delta L =2$ processes~\cite{Atre:2009rg}.
These decays become possible only in the presence of Majorana neutrino mass term and, therefore, don't have direct Standard Model analogues, but the process of their reconstruction, especially in the case of short-lived tau-mesons in the final state, poses an experimental challenge, making studies of electron flavour more promising in the near future.

We present that value described by equation \eqref{eq:2HNL_min} in Figs. \ref{fig:limits_e}, \ref{fig:limits_mu}, \ref{fig:limits_tau} together with the similar results for three HNLs case (see section \ref{sec:3HNL}).

\subsection{Pseudodegenerate state}
%In the previous subsection, we studied the minimal value $U_{min,\alpha}^2$ achievable for mixing with a specific flavor $\alpha$.
In this subsection, we study the relations of mixing of different HNLs with the same flavor.
We also study the mixing of a specific HNL with different flavors in two distinct cases: close to minimal mixing value $\frac{|m_{\alpha\alpha}|}{M}$ and for much greater values of mixing.

\begin{figure}
\centerline{
\includegraphics[width=0.5\textwidth]{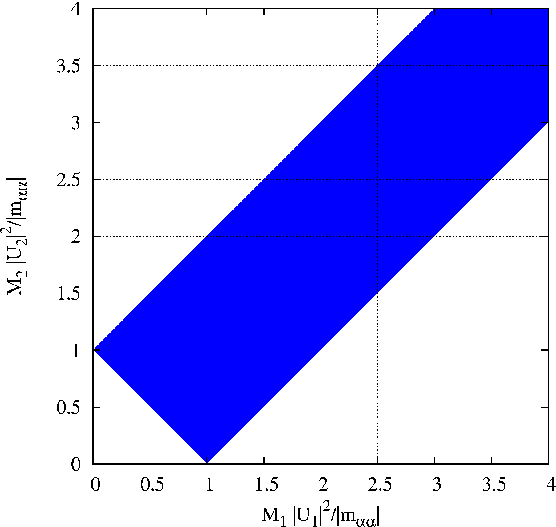}
\hskip 0.05\textwidth
\includegraphics[width=0.5\textwidth]{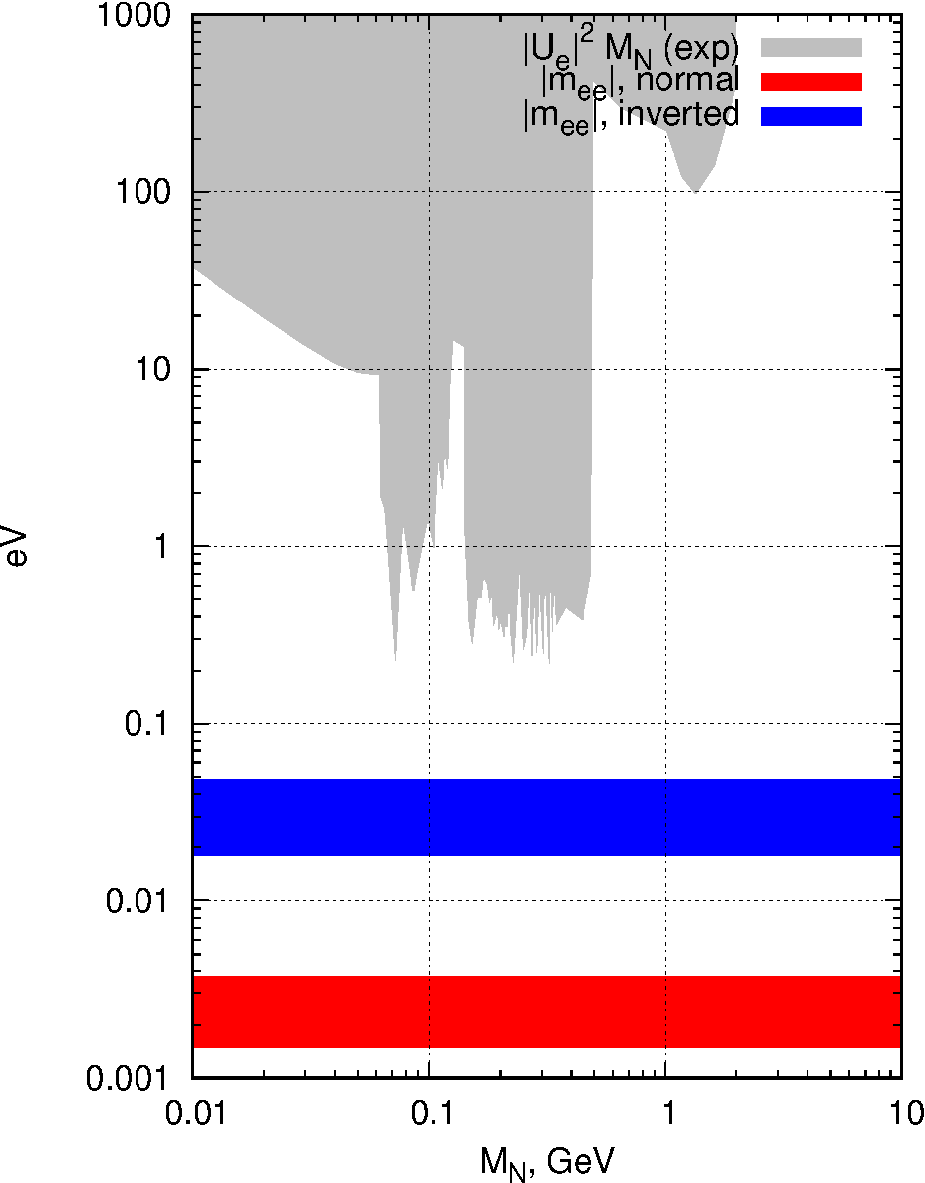}
}
\caption{Left panel: available region for $\mathcal{M}_{1 \alpha}$ and $\mathcal{M}_{2 \alpha}$, where $\alpha \subset \{ e, \mu, \tau \}$. Right panel: current experimental limits on $M_N |U_{eN}|^2$ and  available values of $|m_{ee}|$ (for different $\psi$).}
\label{fig:2HNL}
\end{figure}

It is convenient to introduce notional masses $\mathcal{M}_{i \alpha} = M_i |U_{i \alpha }|^2$.
One can notice that for all $x,y$ (see appendix \ref{app:2HNL_pseudo}):
\begin{align}
\label{eq:notional_close}
|\mathcal{M}_{2 \alpha} - \mathcal{M}_{1 \alpha}| \leq& |m_{\alpha\alpha}|\\
\mathcal{M}_{1 \alpha} + \mathcal{M}_{2 \alpha} \geq& |m_{\alpha\alpha}|
\end{align}
One can see the available region of ($\mathcal{M}_{1 \alpha}, \mathcal{M}_{2 \alpha}$) plane limited by these inequations on the left side of figure \ref{fig:2HNL}.
We can roughly divide the available region in two distinct areas.
The first one is near the minimally allowed values $\mathcal{M}_{1 \alpha} \approx \mathcal{M}_{2 \alpha} \approx |m_{\alpha\alpha}|$.
The second one is when both notional masses are much greater than the effective mass $\mathcal{M}_{1 \alpha} \approx \mathcal{M}_{2 \alpha} \gg |m_{\alpha\alpha}|$.

We start with the latter case which for our purposes we rephrase as:
\begin{equation}
\label{eq:limit2}
\mathcal{M}_{1 \alpha} + \mathcal{M}_{2 \alpha} \gg |m_{\alpha\alpha}| \geq |\mathcal{M}_{2 \alpha} - \mathcal{M}_{1 \alpha}|
\end{equation}

Here, we can notice an important fact.
Our seesaw limit \eqref{eq:2HNL_min} in terms of notional mass simply turns into expression ${\mathcal{M}_{i_{max} \alpha} \geq |m_{\alpha\alpha}|}$. 
We have drawn the current experimental limits, as well as the allowed regions of $|m_{ee}|$ for both hierarchies (and $m_{lightest}=0$) on the right side of figure \ref{fig:2HNL}.
The mixing with electronic neutrino has stricter experimental bounds compared to mixing with two other flavors, so we present only it here.
From this figure, one can conclude that current experimental limits lay at least a magnitude of order higher than the value of $|m_{ee}|$.

One can find that in the limit \eqref{eq:limit2} the expressions $\frac{U_{\alpha}^2}{<U>^2}$ where $\frac{U_{\alpha}^2}{<U>^2} = \frac{U_{1 \alpha}^2}{|U_{1 e}|^2 + |U_{1 \mu}|^2 + |U_{1\tau}|^2} = \frac{U_{2 \alpha}^2}{|U_{2 e}|^2 + |U_{2 \mu}|^2 + |U_{2 \tau}|^2}$.
These expressions depend only on the active neutrino parameters, not depending on any HNL parameter at all.
We provide analytical formulas in appendix \ref{app:2HNL_pseudo}.

Similar, but a bit less strict limit goes into $\mathcal{M}_{1 \alpha} \approx \mathcal{M}_{2 \alpha} \approx |m_{\alpha\alpha}|$ area:
\begin{equation}
\label{eq:limit}
|\mathcal{M}_{2 \alpha} -\mathcal{M}_{1 \alpha}| \ll \mathcal{M}_{1 \alpha} + \mathcal{M}_{2 \alpha}
\end{equation}
The expressions $\frac{U_{\alpha}^2}{<U>^2}$ in that case depend on the additional parameter $\kappa_i = \sqrt{1 - \frac{|m_{ee}|^2}{4\mathcal{M}_{i e}^2}}, 0< \kappa < 1$ (different for each HNL).

%Limit \eqref{eq:limit2} corresponds to $\kappa \to 1$, defining values $\mathcal{M}_{i e}$ much greater than their minimal value.

One can notice that a region described by \eqref{eq:limit2} corresponds to the case $\kappa_i \to 1, \kappa_2 \to 1$ in \eqref{eq:limit}.
Our results for \eqref{eq:limit2} mirror the results obtained in the much stricter ``symmetric limit''~\cite{Drewes:2018gkc}: $M_1=M_2, |U_{\alpha1}|^2=|U_{\alpha2}|^2$, both in pictures and in formulas.
We call our limit \eqref{eq:limit} \emph{pseudodegenerate state}.
Symmetric limit is a specific case of pseudodegenerate state: $\mathcal{M}_{1 e}=\mathcal{M}_{2 e}$. 
%We state that the ratio $U_e:U_\mu:U_\tau$ is defined by the expression \eqref{eq:Ualpha} and is the same for both HNLs in this limit.
We provide expressions for the ratio $U_e:U_\mu:U_\tau$ in terms of active neutrino parameters in appendix \ref{app:2HNL_pseudo}.
In other words, HNLs don't need to be completely degenerate in their mass and mixing at the same time, instead for the mixing ratio to take these forms it is enough to have equal notional masses.
That means that should we find any evidence of HNL mixing with one flavor, we would simultaneously get a hint of the preferable (for two HNLs scenario) mixing range with the other two flavors, as well as line $|U_{2 \alpha}|^2 = \frac{M_1 }{M_2} |U_{1 \alpha}|^2$ as a function of $M_2$ along which we should search for evidence of the other HNL.
On a side note, if one scans values for $\mathcal{M}_{1 e} \ll \mathcal{M}_{2 e}$ and $\mathcal{M}_{1 e} \gg \mathcal{M}_{2 e}$ one obtains ``general case'' areas figuring in literature~\cite{Drewes:2018gkc}.
Because notional masses are restricted by expression \eqref{eq:notional_close}, the areas outside of $\kappa = 1$ allowed area inevitably are defined by the areas $\mathcal{M}_{1 e} \lesssim |m_{ee}|$ and $\mathcal{M}_{2 e} \lesssim |m_{ee}|$, therefore being of little interest for HNLs searches in the near future.
We refer most of the further study of pseudodegenerate state, such as the preferable areas for best fit $\delta$ value and so forth, to the already performed study of symmetrical limit~\cite{Drewes:2022akb}.

On a side note, we warn that our analysis is done at a tree level in terms of corrections to active neutrino masses.
Loop-level corrections can introduce additional restrictions on the theory or outright rule out certain scenarios \cite{Kersten:2007vk}.
The study of such corrections, though, fall outside the scope of this work and can be a subject of future papers.

\begin{figure}[!htb]
\centerline{
\includegraphics[width=0.5\textwidth]{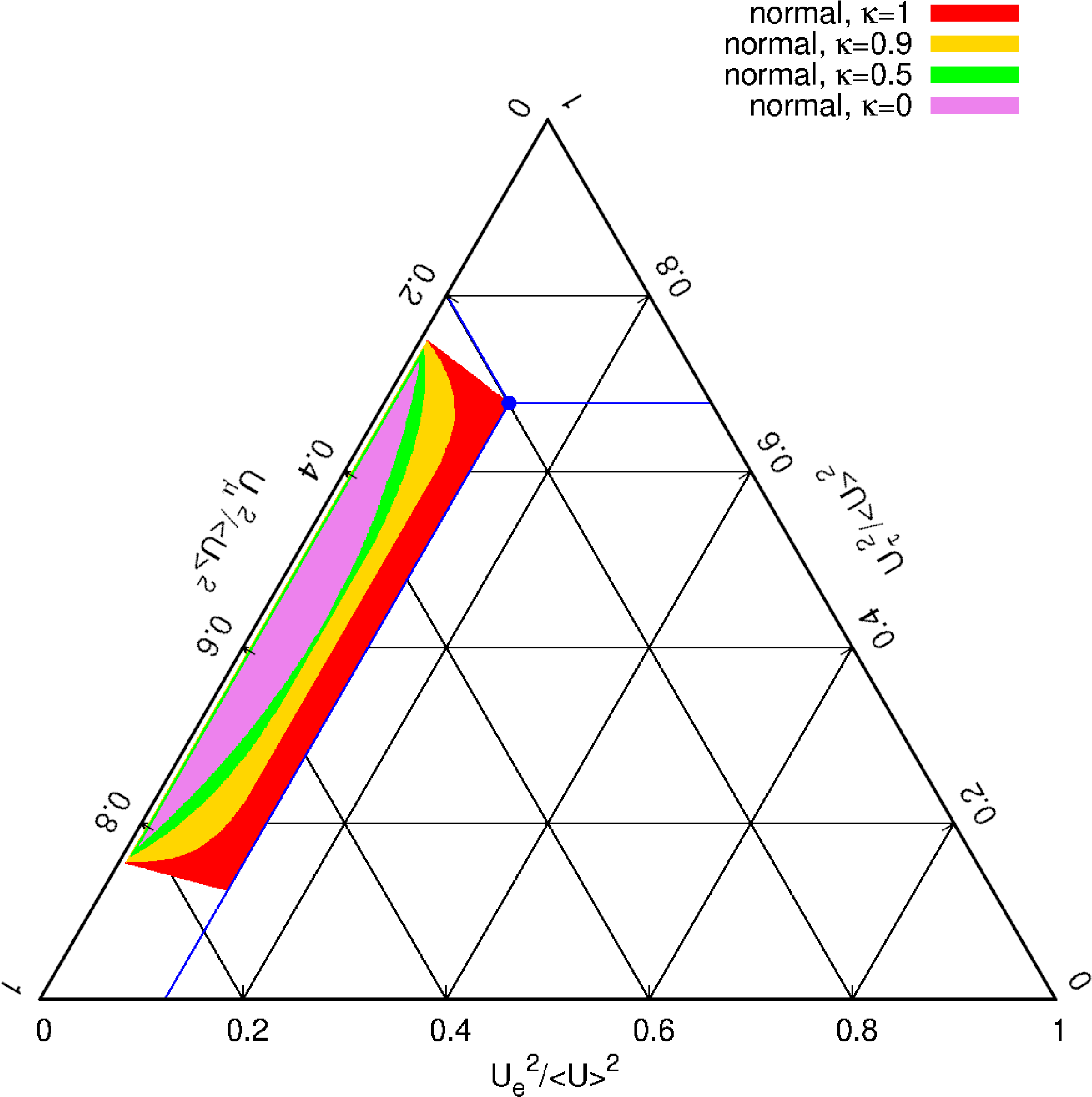}
\includegraphics[width=0.5\textwidth]{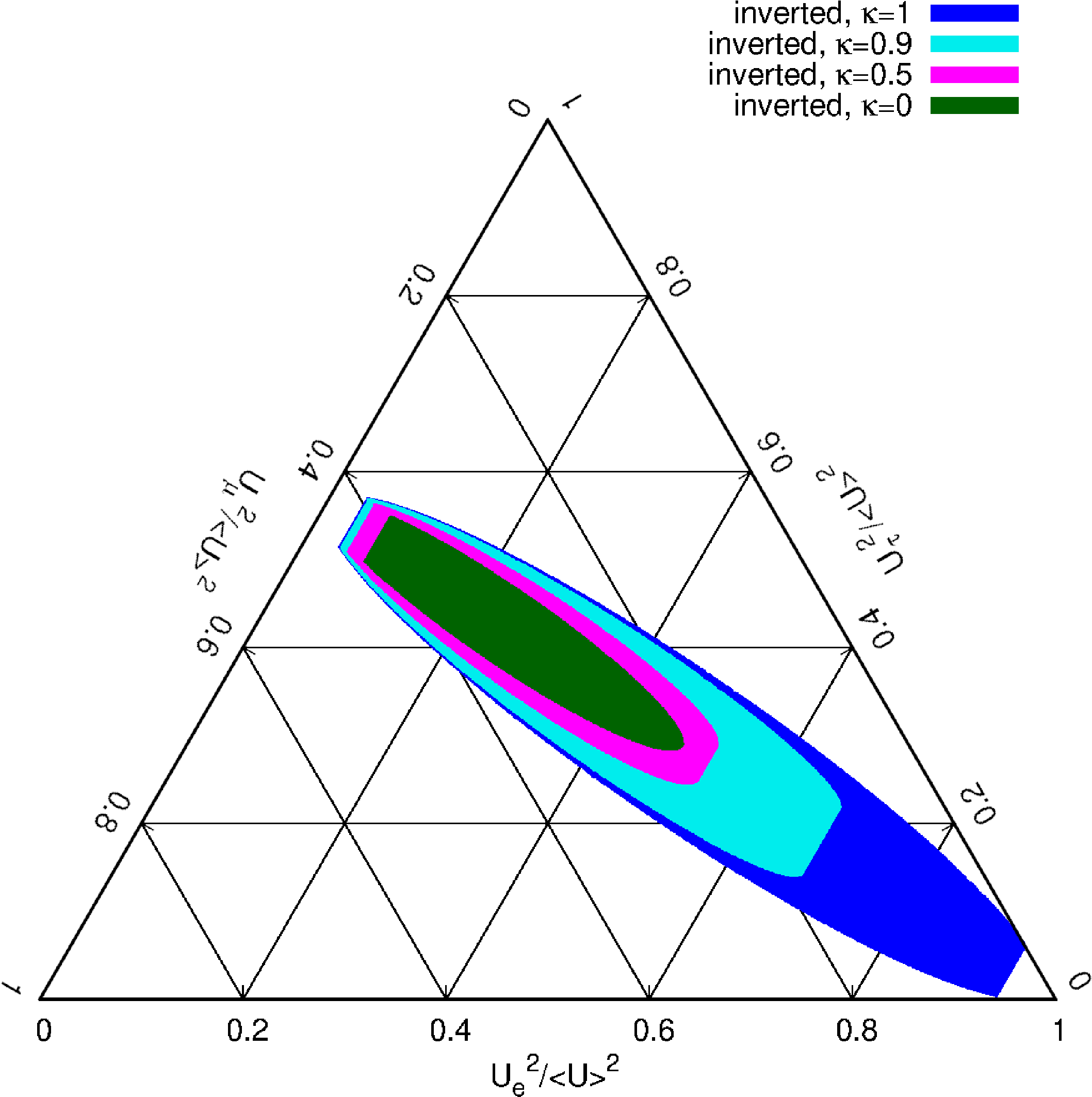}
}
\caption{The available regions ($|U_e|^2,|U_\mu|^2,|U_\tau|^2$) in limit $|\mathcal{M}_{e 2} -\mathcal{M}_{e 1}| \ll \mathcal{M}_{e 1} + \mathcal{M}_{e 2}$. We present results for normal hierarchy on the left panel and for inverted hierarchy on the right panel. Additionally, on the left panel, we present a blue dot (0.12, 0.2, 0.68) with coordinate projection as an example of how to read a specific dot's coordinates on a ``triangular'' grid.}
\label{fig:2HNL_triangular}
\end{figure}

We have drawn the available ratios of mixing in figure \ref{fig:2HNL_triangular}.
%In this figure we use notion $\frac{U_{\alpha}^2}{<U>^2} = \frac{U_{1 \alpha}^2}{U_{1\,tot}^2} = \frac{U_{2 \alpha}^2}{U_{2\,tot}^2}$.
One can see that the allowed regions ``shrink down'' if one decreases the value of $\kappa$.
Therefore, results obtained for limit \eqref{eq:limit} can serve as a conservative boundary for limit \eqref{eq:limit2} as well.
Note, unlike limit \eqref{eq:limit2} two HNLs can have different values of $\kappa$, making the ratio $U_e^2:U_\mu^2:U_\tau^2$ differ between them.
In two HNL case we have $m_{lightest}=0$, and, consequently, $|m_{\alpha\alpha}|>0$.
That means that both HNLs mix with each neutrino flavor, even if mixing with some flavors is greatly suppressed.
We study the behaviour of $|m_{\alpha\alpha}|$ for nonzero values of $m_{lightest}$ in the three-HNLs case.

Additionally, we have checked the dependence of minimal allowed values of $|U_{i\,tot}^2|$ on the mass $M_i$ in the presence of relevant experimental limits.
Unremarkably, the experimental limits at the moment are too weak and don't contribute to the resulting lines in any significant way: they take the shape $|U_{i\,tot}^2|_{min} = \frac{m_j}{M_i}$, where $m_j = m_2$ for normal hierarchy and $m_j = m_1$ for inverted hierarchy.

To summarize our results for the case of two HNLs, we obtained that the minimal value of mixing with a given flavor $\alpha \in \{ e, \mu, \tau \}$ can be expressed as $\frac{|m_{\alpha\alpha}|}{M_{HNL}}$.
The right side of Figure \ref{fig:2HNL} shows us that the minimally allowed values of mixing are, most likely, out of reach of experiments conducted in the near future.
In turn, that means that should any evidence of HNL be found, two HNLs scenario predicts with a high level of precision that $\mathcal{M}_{1 \alpha} = \mathcal{M}_{2 \alpha} \gg |m_{\alpha\alpha}|$ and that ratio of mixing with different flavors is governed by expression provided in appendix \ref{app:2HNL_pseudo} in formula \eqref{eq:Ualpha}, and is the same for both HNLs.
%This would allow one to greatly restrict the search area for mixing of the second HNL with said flavor and the mixing of the first HNL with other flavors.

\section{Three HNLs case}
\label{sec:3HNL}

The equation \eqref{eq:U} holds true for three HNLs case.
The main difference comes from the number of HNL sector parameters.
Matrix $R$ is $3 \times 3$ orthogonal matrix described by three complex angles.
We provide a detailed parametrization and our analytical calculations in appendix \ref{app:3HNL}.
We show there that the lower limit in three HNL case leads to the same equation \eqref{eq:2HNL_min}:
\begin{align}
\label{eq:U_e_min}
U^2_{min, \alpha} =& \frac{|m_{\alpha\alpha}|}{M_{max}},
\end{align}
where $|m_{\alpha\alpha}| = |m_1 U^{\dagger^2}_{{PMNS}_{1\alpha}} e^{i\alpha_1} + m_2 U^{\dagger^2}_{{PMNS}_{2\alpha}} e^{i\alpha_2}+ m_3 U^{\dagger^2}_{{PMNS}_{3\alpha}}|$, $M_{max}=max\{M_1, M_2, M_3\}$. 

\subsection{Lower limit in three HNLs case}
\label{sec:m_alpha_alpha}
Taking into account the result we obtained in eq. \eqref{eq:U_e_min}, the value $|m_{\alpha\alpha}|$ becomes crucial to know the lower bound on see-saw allowed region.
Remember, $|m_{ee}|$ emerges in the neutrinoless double $\beta$ decay~\cite{ParticleDataGroup:2024cfk}.

\begin{figure}[htb]
\includegraphics[width=0.9\textwidth]{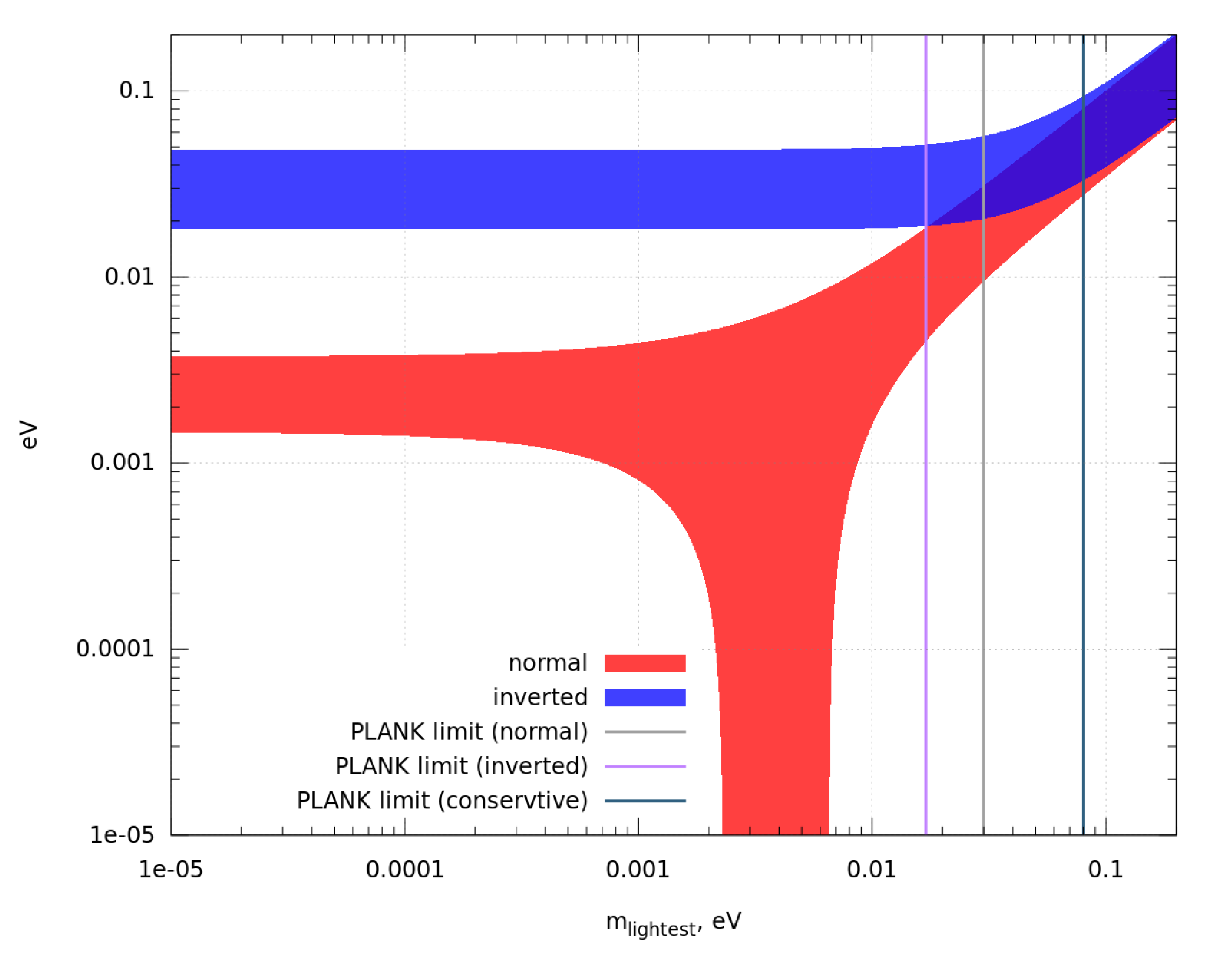}
\caption{The available values of $|m_{ee}|$ as a function of $m_{lightest}$.}
\label{fig:m_ee}
\end{figure}
\begin{figure}[htb]
\includegraphics[width=0.9\textwidth]{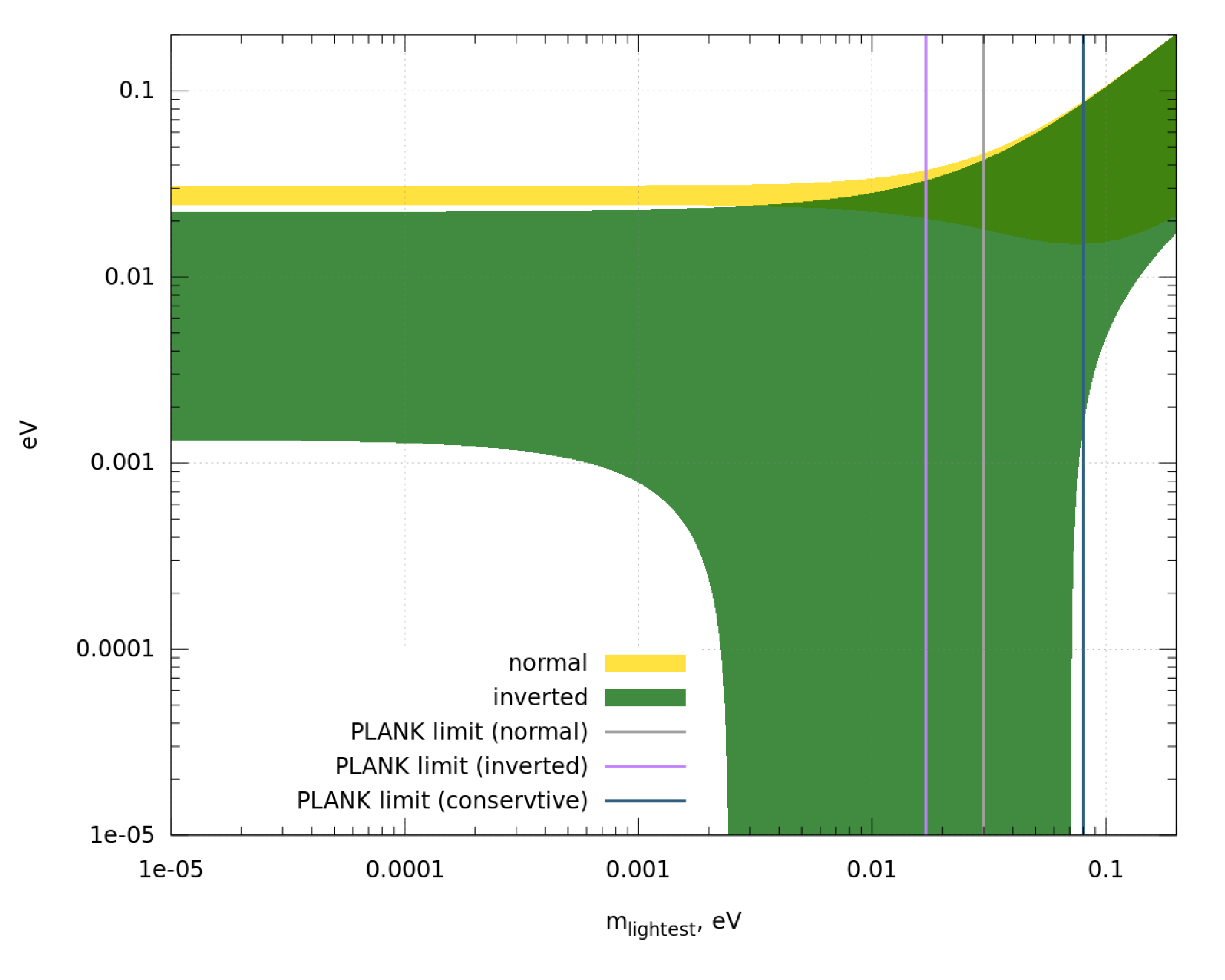}
\caption{The available values of $|m_{\mu\mu}|$  as a function of $m_{lightest}$.}
\label{fig:m_mumu}
\end{figure}
\begin{figure}[htb]
\includegraphics[width=0.9\textwidth]{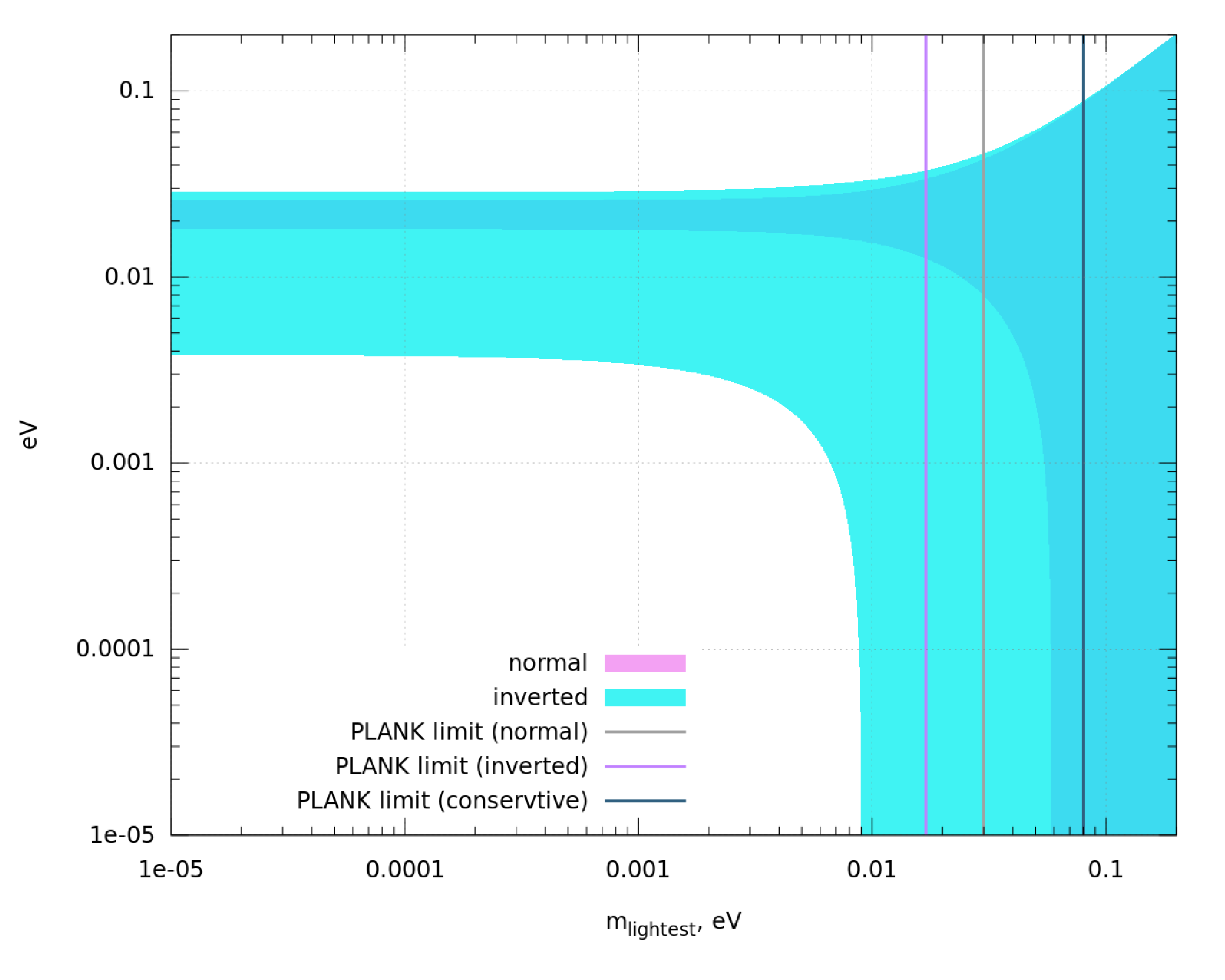}
\caption{The available values of $|m_{\tau\tau}|$ as a function of $m_{lightest}$.}
\label{fig:m_tautau}
\end{figure}

We take central values of active neutrino sector parameters, except for the three CP-violating phases, which we scan from $-\pi$ to $\pi$.
We show in Figures \ref{fig:m_ee}, \ref{fig:m_mumu},  \ref{fig:m_tautau} the resulting allowed regions of $|m_{ee}|, |m_{\mu\mu}|, |m_{\tau\tau}|$ as a function of lightest neutrino mass.
One can see three types of behaviour, depending on the mass of lightest neutrino:
\begin{itemize}
\item ``constant'', $c_1 < |m_{\alpha\alpha}|< c_2$. Limits are approximately unchanging for small values of $m_{lightest}$.
\item ``cancelling''. Allows solution $|m_{\alpha\alpha}|=0$ while the upper limit is usually the transitional stage between constant and linear dependence on $m_{lightest}$.
\item ``linear''. The allowed region is sandwiched between two lines: $a\times m_{lightest} < |m_{\alpha\alpha}|< m_{lightest}$, where $0<a<1$. This is a characteristic behaviour for big values of  $m_{lightest}$, although $m_{\tau\tau}$ experiences cancelling for all values of mass above a certain threshold.
\end{itemize}

In particular, for normal hierarchy $|m_{ee}|$ experiences ``cancelling'' in region 2.3~meV~$< m_{lightest} <$~6.6~meV; $|m_{\mu\mu}|$ doesn't experience ``cancelling'' at all; $|m_{\tau\tau}|$ experiences ``cancelling'' for $m_{lightest} >$~0.59~eV. 
For inverted hierarchy $|m_{ee}|$ doesn't experience ``cancelling''; $|m_{\mu\mu}|$ experiences ``cancelling'' in region 2.4~meV~$< m_{lightest} <$~71~meV; $|m_{\tau\tau}|$ experiences ``cancelling'' for $m_{lightest} >$~90~meV. 

We have some constraints on neutrino masses from laboratory experiments such as KATRIN \cite{KATRIN:2021uub}, that place limits on the expression $m_\nu^2 = \sum_i |U_{ie}|^2 m_i^2 < 0.64$ eV, or, in our terms, $m_{lightest} < 0.8$ eV (the result is the same for both hierarchies in the leading order of magnitude and is practically reduced to $m_{lightest}  \approx \sqrt{m_\nu^2}$).

The sum of masses of active neutrinos plays a significant role in cosmology and therefore can be constrained somewhat by CMB analysis.
In particular, a more conservative PLANCK limit gives us $\sum m_i < 0.26$ eV~\cite{ParticleDataGroup:2024cfk}) which corresponds to $m_{lightest} < 0.08$ eV (the result is the same for both hierarchies in the leading order of magnitude).
Other interpretations of PLANK data can make this limit as low as  $\sum m_i < 0.12$ eV \cite{eBOSS:2020yzd,Planck:2018vyg}, which corresponds to $m_{lightest} < 0.03$ eV for normal hierarchy and $m_{lightest} < 0.017$ eV for inverted hierarchy.
Figures \ref{fig:m_ee}, \ref{fig:m_mumu},  \ref{fig:m_tautau} show that this limit disfavors linear behaviour, but still allows cancelling to occur -- only for normal hierarchy $|m_{\tau\tau}|$ cancelling area becomes disfavored.

One can see that, depending on the unknown value of $m_{lightest}$, our limit \eqref{eq:U_e_min} can take zero value.
That result is most intriguing because it suggests that when such ``cancelling'' occurs, the see-saw mechanism allows the case where no HNL is mixing with a given flavor while mixing with remaining flavors still provides active neutrinos three mass states.
This assessment, though, doesn't consider HNLs contribution to the effective neutrino masses $m_{ee}, m_{\mu\mu}, m_{\tau\tau}$ and other non-leading order corrections to these expressions, which are extensively studied in the literature~\cite{Bezrukov:2005mx,Bolton:2022tds,Schubert:2022lcp}.
The study of the specifics of HNL physics in these `cancelling'' areas falls outside of this work's scope but highlights the importance of the determination of the lightest active neutrino mass and active neutrino mass hierarchy for HNL physics.

We have numerically checked the limits on mixing with a given flavor in ``cancelling'' regions in the presence of current experimental limits on mixing with other flavors to see if such cancelling might only be achieved in the already excluded areas.
We have found that the mixing with one flavor in the ``cancelling'' area can be indistinguishable from zero (we have stopped the computation for $|U_\alpha|^2<10^{-14}$) while mixing with other flavors takes allowed values, even if they can be close to the existing experimental limits.
Therefore, ``cancelling'' areas are an interesting subject for more in-depth study in the future.
\begin{figure}
\includegraphics[width=0.9\textwidth]{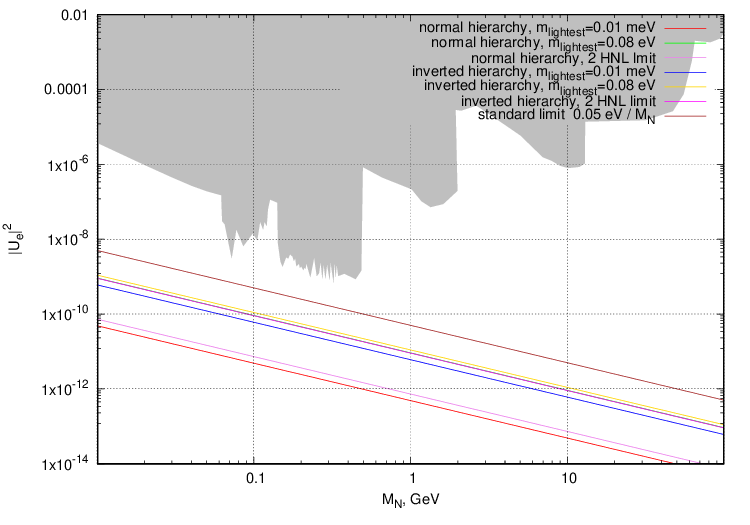}
\caption{Experimental limits and our see-saw lower boundary for different values of $m_{lightest}$ and usually adopted in literature limit $\frac{0.05\textrm{eV}}{M_N}$~\cite{Abdullahi:2022jlv} for mixing with electron neutrino.}
\label{fig:limits_e}
\end{figure}
\begin{figure}
\includegraphics[width=0.9\textwidth]{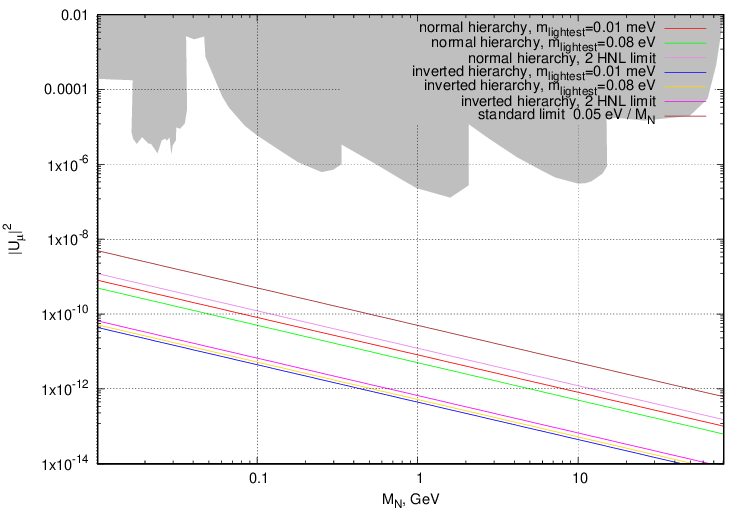}
\caption{Experimental limits and our see-saw lower boundary for different values of $m_{lightest}$ and usually adopted in literature limit $\frac{0.05\textrm{eV}}{M_N}$~\cite{Abdullahi:2022jlv} for mixing with muon neutrino.}
\label{fig:limits_mu}
\end{figure}
\begin{figure}
\includegraphics[width=0.9\textwidth]{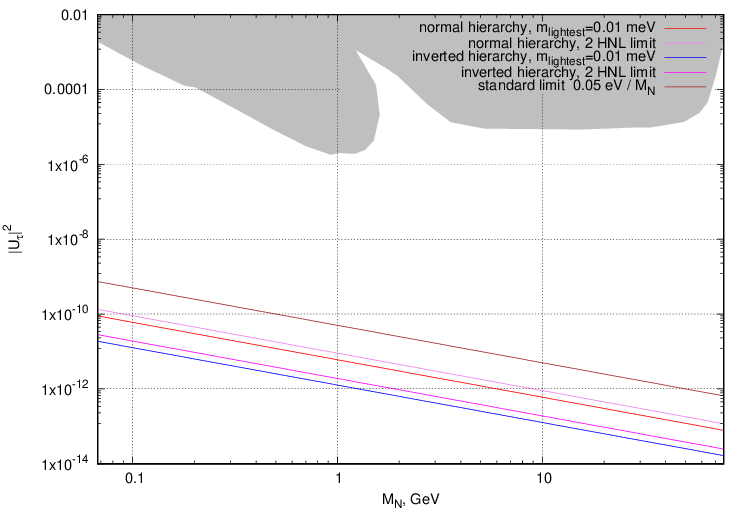}
\caption{Experimental limits and our see-saw lower boundary for different values of $m_{lightest}$ and usually adopted in literature limit $\frac{0.05\textrm{eV}}{M_N}$~\cite{Abdullahi:2022jlv} for mixing with tau neutrino.}
\label{fig:limits_tau}
\end{figure}

In figures \ref{fig:limits_e}, \ref{fig:limits_mu}, \ref{fig:limits_tau} we show the line $\frac{|m_{\alpha\alpha}|}{M_N}$ for several benchmark values of $m_{lightest}$ together with current experimental limits and a usually adopted in literature limit $\frac{0.05\textrm{eV}}{M_N}$~\cite{Abdullahi:2022jlv}.
One can notice that these lines lay lower than previous estimates for the see-saw limit and might significantly depend on the value of $m_{lightest}$ and the hierarchy of active neutrinos.
The determination of the hierarchy of active neutrino masses is expected to be achieved in the coming decade.
These discoveries will be paramount for the determination of a lower boundary of HNLs mixing with active neutrinos in a see-saw model.
If successful, the ongoing neutrinoless double beta decay experiments can help to determine the value of $|m_{ee}|$ or at least significantly restrict it in case of no evidence.
Because ``cancelling'' can't occur to mixing with electron neutrino and mixing with muon neutrino for the same hierarchy and the same value of $m_{lightest}$, it is important to study mixing with these two flavors in conjunction.
Mixing with $\tau$ at present looks less promising: to rule out ``cancelling'' one needs to have a solid restriction on the lightest neutrino mass at level $m_{lightest} < $59~meV for normal hierarchy and $m_{lightest} < $9~meV for inverted hierarchy.
At present, only more optimistic cosmological measures can provide such limits, which are somewhat model-dependent.

\subsection{Pseudodegenerate state}

In three HNLs case the relation between notional masses becomes more complicated.
We provide corresponding analysis in appendix \ref{app:3HNL_pseudo}.
The three effective masses $m_{\alpha\alpha}$ define the minimal values scale according to the equation \eqref{eq:U_e_min}, yet there are three notional masses tied to each other:

\begin{align}
\label{eq:23}
|\mathcal{M}_{3 \alpha} - &\mathcal{M}_{2 \alpha}| - |m_{\alpha\alpha}| \leq   \mathcal{M}_{1 \alpha} \, \leq \, \mathcal{M}_{2 \alpha} + \mathcal{M}_{3 \alpha} + |m_{\alpha\alpha}|\\
|\mathcal{M}_{3 \alpha} - &\mathcal{M}_{1 \alpha}| - |m_{\alpha\alpha}| \leq \mathcal{M}_{2 \alpha} \, \leq \, \mathcal{M}_{1 \alpha} + \mathcal{M}_{3 \alpha} + |m_{\alpha\alpha}|\\
|\mathcal{M}_{2 \alpha} - &\mathcal{M}_{1 \alpha}| - |m_{\alpha\alpha}| \leq \mathcal{M}_{3 \alpha} \, \leq \, \mathcal{M}_{1 \alpha} + \mathcal{M}_{2 \alpha} + |m_{\alpha\alpha}|
\end{align}

\begin{figure}
\includegraphics[width=0.9\textwidth]{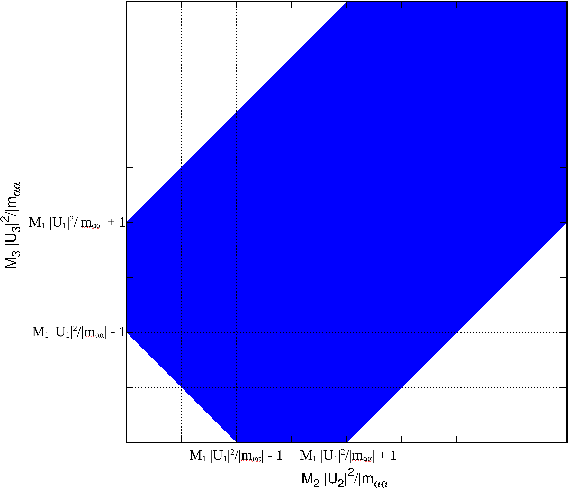}
\caption{Available region for $M_2 |U_{\alpha 2}|^2$ and $M_3 |U_{\alpha 3}|^2$, where $\alpha \subset \{ e, \mu, \tau \}$.}
\label{fig:3HNL_u}
\end{figure}

We present the available region of ($\mathcal{M}_{2 \alpha}, \mathcal{M}_{3 \alpha}$) plane limited by these inequations for a fixed value of $\mathcal{M}_{1 \alpha}$ in figure \ref{fig:3HNL_u}.
Obviously, if $\mathcal{M}_{1 \alpha} < |m_{\alpha\alpha}|$ that would mean that $\mathcal{M}_{3 \alpha}\geq 0$ for $\mathcal{M}_{2 \alpha} < \mathcal{M}_{1 \alpha} + |m_{\alpha\alpha}|$ (as it can't take negative values).

We are interested in the following limit:
\begin{equation}
\label{eq:3HNL_limit}
|\mathcal{M}_{3 \alpha}- \mathcal{M}_{2 \alpha}| \ll \mathcal{M}_{2 \alpha} + \mathcal{M}_{3 \alpha},
\end{equation}
that allows us to say that $\mathcal{M}_{2 \alpha} \approx \mathcal{M}_{3 \alpha} \approx \frac{1}{2} \left( \mathcal{M}_{2 \alpha} + \mathcal{M}_{3 \alpha}\right)$.
From the experimental point of view, we are once again mostly interested in area $|m_{\alpha\alpha}| \ll \mathcal{M}_{i \alpha}$, which leads us to two distinct limits when pseudodegenerate state is achieved:
\begin{align}
\label{eq:3HNL_requirements}
\mathcal{M}_{1 \alpha} \ll \mathcal{M}_{3 \alpha} + \mathcal{M}_{2 \alpha}, & |m_{\alpha\alpha}| \ll \mathcal{M}_{2 \alpha} + \mathcal{M}_{3 \alpha}\\
\label{eq:same_magnitude_massmixing}
\mathcal{M}_{1 \alpha} \gg |\mathcal{M}_{3 \alpha} - \mathcal{M}_{2 \alpha}|, & |m_{\alpha\alpha}| \ll \mathcal{M}_{2 \alpha} + \mathcal{M}_{3 \alpha}
\end{align}
In these formulas, we haven't fixed the hierarchy of notional masses in any way.
As we can see from eq. \eqref{eq:3HNL_requirements}, pseudodegenerate state is automatically achieved when one notional mass $\mathcal{M}_{i \alpha}$ is much less than the other two.
When all three parameters are of the same magnitude $\mathcal{M}_{1 \alpha}~\sim~\mathcal{M}_{2 \alpha}~\sim~\mathcal{M}_{3 \alpha}$ we generally don't have pseudodegenerate state, but it can still be achieved by satisfying the criteria \eqref{eq:same_magnitude_massmixing}.
It makes possible to achieve the pseudodegenerate state for two smaller notional masses.
Curiously, according to \eqref{eq:23}, the bigger notional mass can't be greater than the sum of the other two, inevitably still placing all three notional masses at the same scale of magnitude in such a scenario.
In other words, it is possible to have HNL mass hierarchy with one small mass and two big masses, but not the other way around.

In the pseudodegenerate state, we can once again study the ratio $|U_e|^2:|U_\mu|^2:|U_\tau|^2$ that won't depend on HNLs masses.
Simplifying the expressions, we can obtain similar formulas and graphs to the ones we have for two HNLs case.
In limit \eqref{eq:3HNL_limit} this ratio is the same for mixing with the second and the third HNL.
The resulting expression is more complex in three HNLs case compared to two HNLs case, as it depends not only on the active sector's parameters, but also on two complex sector angles from HNLs sector.
We provide corresponding expressions in  appendix \ref{app:3HNL_pseudo}.

One can notice that $\frac{|U_{i \alpha}|^2}{|U_{i\,tot}|^2} = 0$ is only possible when $|m_{\alpha\alpha}|=0$, which, as we have studied, can be achieved for some values of the mass of the lightest active neutrino.

\begin{figure}[htb]
\centerline{
\includegraphics[width=0.5\textwidth]{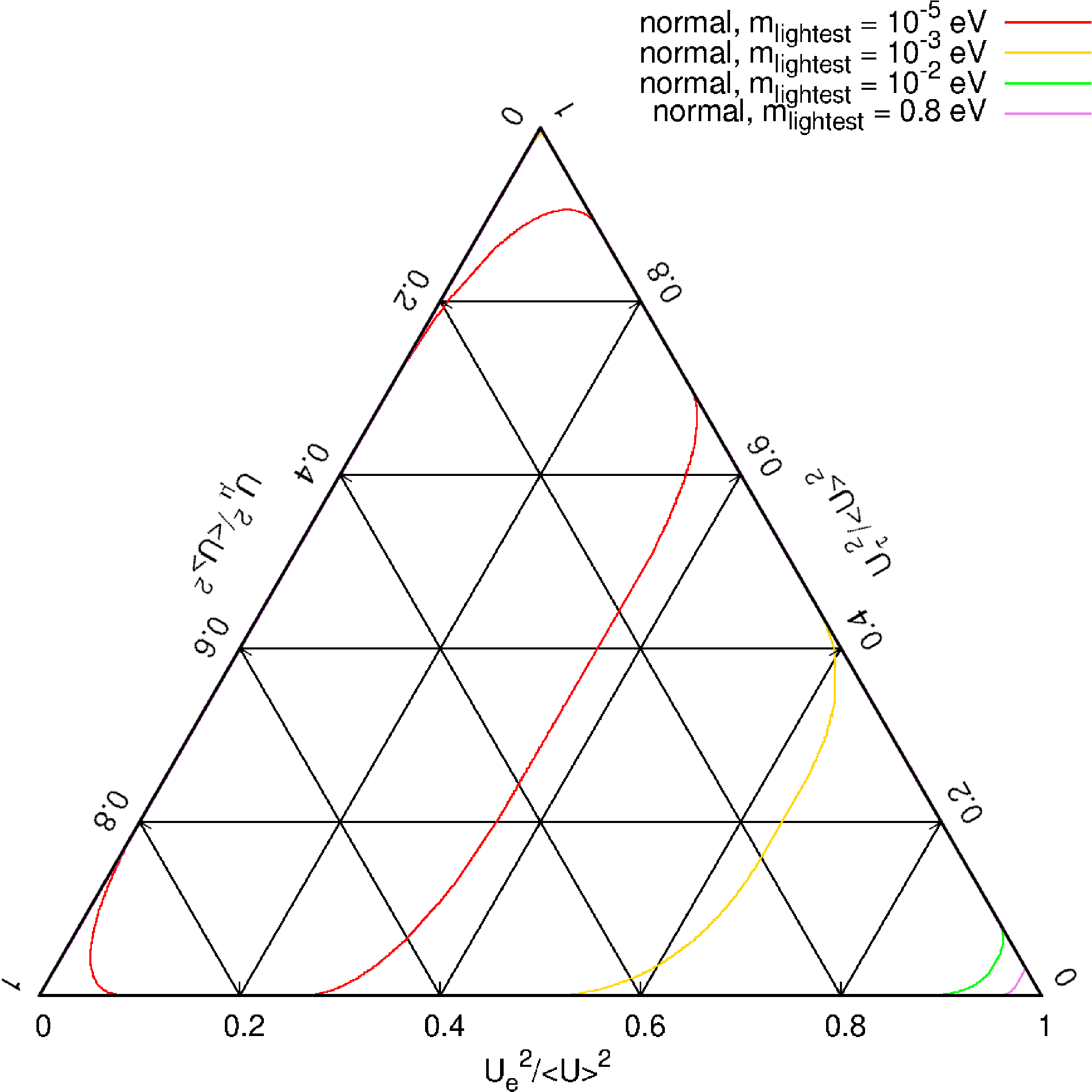}
\includegraphics[width=0.5\textwidth]{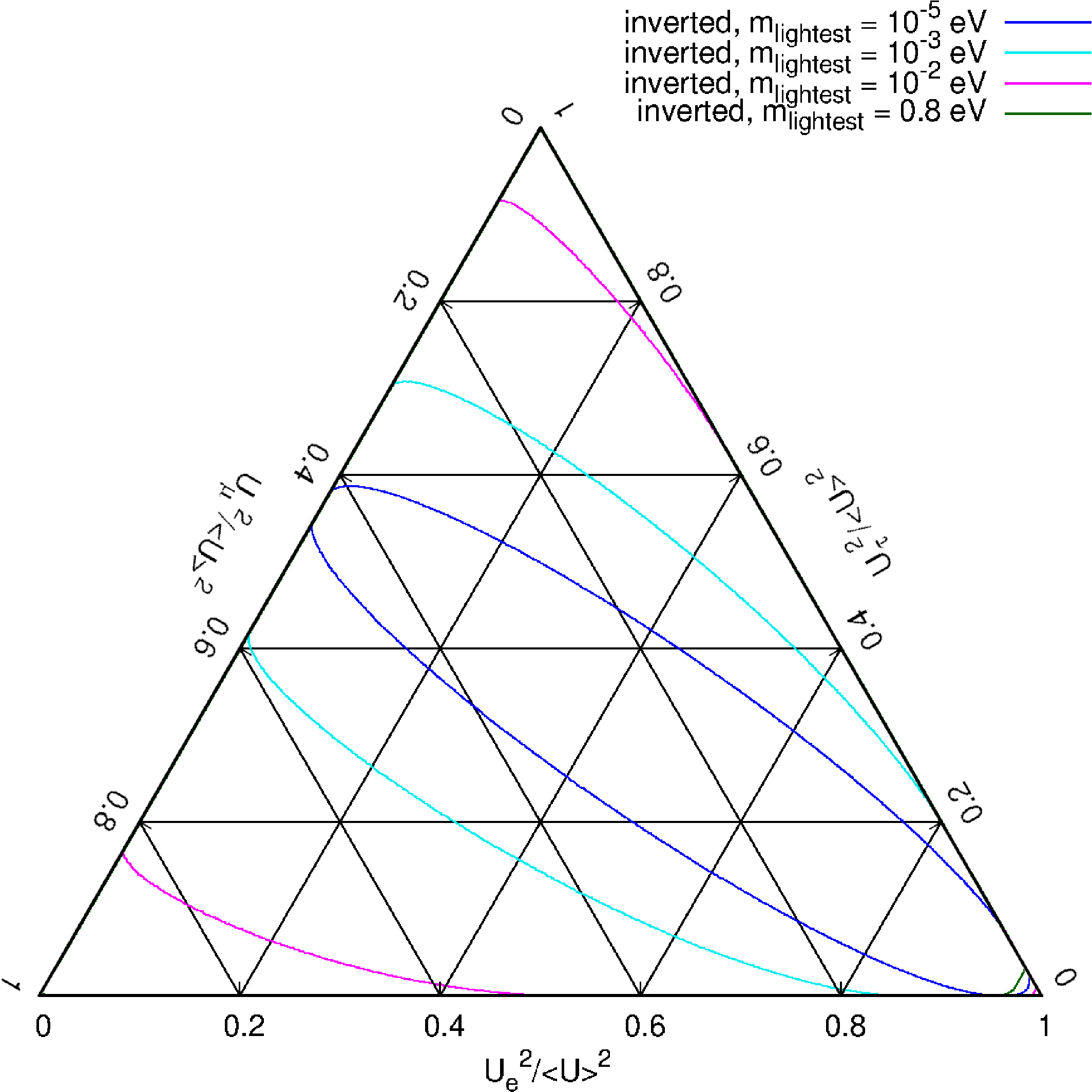}
}
\caption{The available regions ($|U_e|^2,|U_\mu|^2,|U_\tau|^2$) in limit \eqref{eq:3HNL_requirements} -- \eqref{eq:3HNL_limit}. Left panel: normal hierarchy, right panel: inverted hierarchy.}
\label{fig:3HNL_mixing}
\end{figure}

We present the resulting mixing in figure~\ref{fig:3HNL_mixing}.
Our results are consistent with similar study~\cite{Chrzaszcz:2019inj}.
Once again, we obtain our results for a less strict pseudodegenerate state limit \eqref{eq:3HNL_limit}.
In terms of available ratio $|U_e|^2:|U_\mu|^2:|U_\tau|^2$, one can see that for larger values of $m_{lightest}$, almost all parameter space is available.
For example, for $m_{lightest} = 0.08$ eV only a small area near $|U_e|^2 \approx 1$ remains forbidden.
Therefore, from the perspective of searches for HNLs, the question of hierarchy and the value of $m_{lightest}$ remain, maybe, the most crucial factor, as they can greatly affect the available mixing ratio $|U_e|^2:|U_\mu|^2:|U_\tau|^2$.

To summarize, we obtained similar results for the case of three HNLs to those we had for the case of two HNLs.
The main difference for the lower boundary lies in its dependence on unknown value $m_{lightest}$.
In special cases, the naive approach even allows the absence of mixing with a given flavor for all HNLs.
Therefore at present, no conservative lower bound can be placed for any mixing.
That can change, however, with the determination of active neutrino mass hierarchy: at that point either mixing with $e$ or $\mu$ would have a solid boundary $\frac{|m_{\alpha\alpha}|}{M_{HNL}}$, reaching which would allow to rule out see-saw mechanism in a studied HNL mass region.
That promotes the importance of simultaneous experimental studies of mixing of HNLs with both of these flavors.
At the same time, this boundary lay significantly lower than existing experimental limits.
We find that pseudegenerate state is usually achieved when one HNL is effectively decoupled from the other two, but our results show that even so the resulting ratios for mixing with different flavors greatly differ from the one we have in two HNLs case.
Overall, three HNLs case is significantly less restricted compared to the two HNLs case, but can still provide viable hints in future HNL searches.

\section{Conclusion}
\label{sec:conclusions}
In this work, we have studied the see-saw limits on the mixing of heavy neutral leptons with active neutrinos in the case of two and three such HNLs.
We obtained the minimal mixing value with any given flavor for both of these cases and presented the available area for expressions $|m_{ee}|, |m_{\mu\mu}|, |m_{\tau\tau}|$, closely related to them.
Coincidentally, value $|m_{ee}|$ turned out to be an effective neutrino mass that appears in neutrinoless double beta decay searches.
In three HNLs case we show that these expressions can become zero depending on active neutrino parameters, calling for a more in-depth study of next-order contributions that can affect said mixing.
Moreover, we show that currently adopted see-saw limits lay significantly higher than our most conservative estimates.
Unfortunately, three HNLs mixing limit predictions are similar to the two HNLs predictions but greatly depend on the value of $m_{lightest}$.
As such, future experiments that can determine active neutrino mass hierarchy and further limit the value of $m_{lightest}$ will greatly reduce the uncertainty for several aspects of HNL searches.
We show that if we are to find an HNL signal in the near future, assuming two HNLs case, it would mean that HNL parameters are realised in a pseudodegenerate state.
This state is also realised for most of the usually studied variations of three HNLs case, when one HNL effectively decouples from the other two.
For this state, the ratio of mixing with different flavors is greatly restricted,.
These restrictions are stricter for two HNLs case compared to three HNLs case.
We show that this restriction criterion coincidences with the criterion already studied in the symmetrical limit \cite{Abdullahi:2022jlv,Drewes:2018gkc,Chrzaszcz:2019inj}.
From our point of view, it a specific case of pseudodegenerate state, more favored from a theoretical point of view than our generalization.
We proclaim that most of the results obtained in symmetrical limit can be extended to the pseudodegenerate state.

\bmhead{Acknowledgments}
We would like to express gratitude to Dmitry Gorbunov and Yury Kudenko for the valuable discussions and suggestions we had during this manuscript's preparation.

\begin{appendices}

\section{Active sector parametrization}
\label{sec:appendix}

Experiments provide us with differences in the squared mass of active neutrinos \cite{ParticleDataGroup:2024cfk}:
\begin{equation}
\label{data2}
\begin{array}{cccc}
&& best\,fit & 3\sigma\,range\\
\Delta m_{21}^2 \big[ 10^{-5} \, \textrm{eV}^2\big] & = &  7.39, & 6.79 - 8.01\\
|\Delta m_{32}^2| \big[ 10^{-3} \, \textrm{eV}^2\big] & = &  2.449, & 2.358 - 2.544\\
&& (2.509), & (2.416 - 2.603)
\end{array}\nonumber
\end{equation}
where, $\Delta m_{21}^2 = m_2^2 - m_1^2$ and $\Delta m_{32}^2 = m_3^2 - m_2^2$.
It is usually defined that $m_2>m_1$ (just for convenience), but we don't know which of $m_1, m_3$ is smaller.
This results in two viable scenarios for neutrino mass hierarchy: \emph{normal} ($m_1<m_2<m_3$) and \emph{inverted} ($m_3<m_1<m_2$).
To date, there is no viable method to determine the mass of the lightest neutrino $m_{lightest}$, yet neutrinoless double beta decay searches~\cite{ParticleDataGroup:2024cfk} and cosmology~\cite{ParticleDataGroup:2024cfk,eBOSS:2020yzd,Planck:2018vyg} can provide some insights.

For the normal hierarchy, we have:
\begin{displaymath}
\begin{array}{rcl}
m_1 & = & m_{lightest}\\
m_2 & = & \sqrt{m_{lightest}^2 + \Delta m_{21}^2}\\
m_3 & = & \sqrt{m_{lightest}^2 + \Delta m_{21}^2  + |\Delta m_{32}^2|},
\end{array}
\end{displaymath}
and for the inverted hierarchy:
\begin{displaymath}
\begin{array}{rcl}
m_3 & = & m_{lightest}\\
m_1 & = & \sqrt{m_{lightest}^2 - \Delta m_{21}^2  + |\Delta m_{32}^2| }\\
m_2 & = & \sqrt{m_{lightest}^2 + |\Delta m_{32}^2| }.
\end{array}
\end{displaymath}

Of utmost importance is the relation of flavor basis to mass basis, which is described using Pontecorvo-Maki-Nakagawa-Sakata matrix $U_{PMNS}$ multiplied by Majorana phase matrix:
\begin{equation}
\label{eq2}
\left( \begin{array}{ccc}
\nu_1\\
\nu_2\\
\nu_3\\
\end{array} \right)
= \left(\begin{array}{ccc}
e^{i \frac{\alpha_1}{2}} & 0 & 0\\
0 & e^{i \frac{\alpha_2}{2}} & 0\\
0 & 0 & 1\\
\end{array}\right) U_{PMNS}^\dagger \left(\begin{array}{ccc}
\nu_e\\
\nu_\mu\\
\nu_\tau\\
\end{array}\right),
\end{equation}

Majorana CP-violating phases $\alpha_1, \alpha_2$, while being vital part of most theories with heavy neutral leptons, so far have yet to be proven to exist, let alone be limited in any way.
$U_{PMNS}$ matrix is described by three angles $\theta_{12}, \theta_{13}, \theta_{23}$ and Dirac CP-violaing phase $\delta$.

The matrix takes form~\cite{ParticleDataGroup:2024cfk}:
\begin{align}
\label{eq3}
&U_{PMNS}^\dagger= \left(\begin{array}{ccc}
c_{13} c_{12} & -c_{23} s_{12} - s_{23} s_{13} c_{12} e^{-i \delta} & s_{23} s_{12} - c_{23} s_{13} c_{12} e^{-i \delta}\\
c_{13} s_{12} & c_{23} c_{12} - s_{23} s_{13} s_{12} e^{-i \delta} & -s_{23} c_{12} - c_{23} s_{13} s_{12} e^{-i \delta}\\
s_{13}e^{i \delta} & s_{23} c_{13} & c_{23} c_{13}\\
\end{array}\right),
\end{align} 
here $c_{ij} = \cos\theta_{ij}$ and $s_{ij} = \sin\theta_{ij}$, with $i,j=1,2,3, i<j$.

The currently adopted values of these parameters~\cite{ParticleDataGroup:2024cfk}):
\begin{displaymath}
\begin{array}{cccc}
\sin^2 \theta_{12} & = & 0.310, & 0.275  - 0.350\\
\sin^2 \theta_{23} & = & 0.558, & 0.427 - 0.609\\
&& (0.563), & (0.430 - 0.612)\\
\sin^2 \theta_{13} & = & 0.02241, & 0.02046 - 0.02440\\
&& (0.02261), & (0.02066 -0.02461). 
\end{array}
\end{displaymath}

At times, it is more convenient to use angles $\theta_{ij}$ themselves:
\begin{equation}
\label{data1}
\begin{array}{cccc}
\theta_{12} & = & 33.82^\circ & \\
\theta_{23} & = & 48.3^\circ & (48.6^\circ)\\
\theta_{13} & = & 8.61^\circ & (8.65^\circ)
\end{array}\nonumber
\end{equation}

There is still little evidence to prefer a certain value of $\delta$, although there are some hints \cite{ParticleDataGroup:2024cfk}.
In this paper, as a general rule, we treat all three phases as free parameters, $\delta \in [0,2 \pi), \alpha_1 \in [- \pi, \pi), \alpha_2 \in [- \pi, \pi)$.

\section{Heavy neutral leptons parametrization}
\label{app:HNL}
\subsection{Two HNLs case}
\label{app:2HNL}
The neutrino mass hierarchy greatly affects two HNLs case, as it determines which elements of $U_{PMNS}$ matrix play a role in mixing with HNLs.
%We provide more information on active neutrino sector parameters used in this work in the appendix \ref{sec:appendix}.

Matrix $R$ takes slightly different forms depending on the hierarchy:
\begin{align}
\textrm{normal: } m_1=0 \Rightarrow R= \left( \begin{array}{ccc}
0 & c & - \xi s\\
0 & s & \xi c\\
\end{array}\right);\\
\textrm{inverted: } m_3=0 \Rightarrow R= \left( \begin{array}{ccc}
c & - \xi s & 0\\
s & \xi c & 0\\
\end{array}\right),
\end{align}
where $\xi = \pm 1, c =\cos \omega, s=\sin \omega$, and $\omega = x + i y \subset \mathbb{C}$ is not restricted in any way.

The resulting mixing matrix is:
\begin{equation}
\label{eq:U_normal}
U =  i  \times \left( \begin{array}{ccc}
\frac{1}{\sqrt{M_1}}\Gamma_{11} & \frac{1}{\sqrt{M_1}} \Gamma_{12} & \frac{1}{\sqrt{M_1}} \Gamma_{13}\\
\frac{1}{\sqrt{M_2}} \Gamma_{21} & \frac{1}{\sqrt{M_2}} \Gamma_{22} & \frac{1}{\sqrt{M_2}} \Gamma_{23}
\end{array}\right),
\end{equation}
where
\begin{align}
\Gamma_{1i}  \equiv & \lambda_{1i} c - \lambda_{2i} e^{i\psi} s\\
\Gamma_{2i}  \equiv & \lambda_{1i} s + \lambda_{2i} e^{i\psi} c
%A_{ij} & \equiv & U_{PMNS_{ij}}^\dagger|_{\alpha_1=0,\alpha_2=0}, i,j =1, 2, 3
\end{align}
For normal hierarchy ($m_1=0$):
\begin{align}
\lambda_{1i} \equiv \sqrt{m_2} U^\dagger_{{PMNS}_{2i}}; \quad \lambda_{2i} \equiv \sqrt{m_3} U^\dagger_{{PMNS}_{3i}}; \quad e^{i \psi }= \pm e^{-i\frac{\alpha_2}{2}}\nonumber
\end{align}
For inverted hierarchy ($m_3=0$):
\begin{align}
\lambda_{1i} \equiv \sqrt{m_1} U^\dagger_{{PMNS}_{1i}}; \quad \lambda_{2i} \equiv \sqrt{m_2} U^\dagger_{{PMNS}_{2i}}; \quad e^{i \psi }= \pm e^{i\frac{\alpha_2-\alpha_1}{2}}\nonumber
\end{align}
Phenomenologically, each HNL is usually described with its mass $M_I$ and mixing to three neutrino flavors $|U_{I\alpha}|^2$. 
Matrix $U$ can be multiplied by any complex phase without changing these expressions.
In this way only one independent Majorana angle $\psi$ remains, as described above, and multiplier $i$ can be safely omitted.

Notice that:
\begin{align}
\textrm{normal:} |\Gamma_{1\alpha}^2+\Gamma_{2\alpha}^2| &= |\lambda_{1\alpha}^2+\lambda_{2\alpha}^2 e^{2 i \psi}| =|m_2 U^{\dagger^2}_{{PMNS}_{2\alpha}} + m_3 U^{\dagger^2}_{{PMNS}_{3\alpha}} e^{-i\alpha_2} |\nonumber\\
\textrm{inverted:} |\Gamma_{1\alpha}^2+\Gamma_{2\alpha}^2| &= |\lambda_{1\alpha}^2+\lambda_{2\alpha}^2 e^{2 i \psi}| =|m_1 U^{\dagger^2}_{{PMNS}_{1\alpha}} + m_2 U^{\dagger^2}_{{PMNS}_{2\alpha}} e^{i(\alpha_2-\alpha_1)}|\nonumber
\end{align}
The last expression doesn't depend on $\omega$ and serve as a definition of the expression $|m_{\alpha\alpha}| = |m_1 U^{\dagger^2}_{{PMNS}_{1\alpha}} e^{i\alpha_1} + m_2 U^{\dagger^2}_{{PMNS}_{2\alpha}} e^{i\alpha_2} + m_3 U^{\dagger^2}_{{PMNS}_{3\alpha}}|$ for both hierarchies.

We study the minimally available values of an expression $U^2_{sum, \alpha} = |U_{1\alpha}|^2 + |U_{2\alpha}|^2$ that would still allow for measured active neutrino parameters.
The extremum criterion is:
\begin{equation}
\frac{\partial U^2_{sum, \alpha}}{\partial \omega}  = - \frac{1}{M_1} \Gamma_{1\alpha}^*\Gamma_{2\alpha} +  \frac{1}{M_2} \Gamma_{2\alpha}^*\Gamma_{1\alpha} = 0
\end{equation}
This equation can have three solutions:
\begin{align}
\Gamma_{1\alpha}  = & 0\\
\Gamma_{2\alpha}  = & 0\\
M_1  = & M_2, \Gamma_{1\alpha}^*\Gamma_{2\alpha} - \Gamma_{1\alpha}\Gamma_{2\alpha}^* = 0 
\end{align}

In the last equation we only consider $|\Gamma_{1\alpha}|~\neq~0, |\Gamma_{2\alpha}|~\neq~0$.

%Our extremum solutions are scaled by expression $|m_{\alpha\alpha}|$.
When $\Gamma_{1\alpha}=0$ it automatically means that $|\Gamma_{2\alpha}|^2 = |\Gamma_{1\alpha}^2 + \Gamma_{21\alpha}^2| = |m_{\alpha\alpha}|$.
When $\Gamma_{2\alpha}=0$ it automatically means that $|\Gamma_{1\alpha}|^2 = |\Gamma_{1\alpha}^2 + \Gamma_{2\alpha}^2| = |m_{\alpha\alpha}|$.
Therefore, the first two cases right away give us the answer for the minimal value for the case $M_1 \neq M_2$: $U^2_{min, \alpha} = \frac{|m_{\alpha\alpha}|}{M_{max}}$, where $M_{max}=max\{M_1, M_2\}$.

We check that for degenerate mass the result is the same. 
If $M_1 = M_2 \equiv M$ the summary mixing is:
\begin{align}
U^2_{sum, \alpha} =& \frac{1}{M} \left(|\Gamma_{1\alpha}|^2 + |\Gamma_{2\alpha}|^2\right) = \frac{1}{M}  \big((|\lambda_{1\alpha}|^2 + |\lambda_{2\alpha}|^2) \cosh(2y) + \nonumber\\
&\sinh(2 y) 2 \Im \left[ \lambda_{1\alpha}^* \lambda_{2\alpha} e^{i\psi} \right]\big) = \frac{1}{M} \cosh(2(y-y_\alpha)) \times \nonumber\\
&\sqrt{(|\lambda_{1\alpha}|^2 + |\lambda_{2\alpha}|^2)^2 -4 (\Im \left[ \lambda_{1\alpha}^* \lambda_{2\alpha} e^{i\psi} \right])^2 } = \nonumber\\
& = \frac{|m_{\alpha\alpha}|}{M}\cosh(2(y-y_\alpha))
\end{align}
Here $U_{sum, \alpha}^2$ doesn't depend on $x$ anymore and ``$y_\alpha$'' corresponds to the minimal value of $U^2_{sum, \alpha}$, providing us with the same result:
\begin{align}
U^2_{min, \alpha} =&  \frac{1}{M_{max}} |m_1 U^{\dagger^2}_{{PMNS}_{1\alpha}} e^{i\alpha_1} + m_2 U^{\dagger^2}_{{PMNS}_{2\alpha}} e^{i\alpha_2}+ m_3 U^{\dagger^2}_{{PMNS}_{3\alpha}}| \equiv \frac{|m_{\alpha\alpha}|}{M_{max}}
\end{align}

\subsection{Pseudodegenerate state for two HNLs}
\label{app:2HNL_pseudo}
We introduced the notional masses as $\mathcal{M}_{i \alpha} = M_i |U_{i \alpha }|^2$.
From one point of view, these expressions only depend on HNL mixing and mass and potentially can be restricted experimentally.
From other point of view, $\mathcal{M}_{i e} = |\Gamma_{i 1}|^2$, $\mathcal{M}_{i \mu} = |\Gamma_{i 2}|^2$, $\mathcal{M}_{i \tau} = |\Gamma_{i 3}|^2$, depend only on $\omega, \psi, \delta$ and known active neutrino parameters.
These values are closely connected:
\begin{align}
\label{eq:2HNL_y}
\mathcal{M}_{1 \alpha} +  \mathcal{M}_{2 \alpha} =& |\Gamma_{1 \alpha}|^2 + |\Gamma_{2 \alpha}|^2 =(|\lambda_{1 \alpha}|^2 + |\lambda_{2 \alpha}|^2) \cosh(2y) +\nonumber\\
+& 2 \Im \left[ \lambda_{1 \alpha}^* \lambda_{2 \alpha} e^{i\psi} \right] \sinh(2 y) = |m_{\alpha\alpha}| \cosh (2 (y - y_\alpha)),\\
\mathcal{M}_{2 \alpha} - \mathcal{M}_{1 \alpha} =& |\Gamma_{1 \alpha}|^2 - |\Gamma_{2 \alpha}|^2 = (|\lambda_{1 \alpha}|^2 - |\lambda_{2 \alpha}|^2) \cos(2x) - \nonumber\\
-& 2 \Re \left[ \lambda_{1 \alpha}^* \lambda_{2 \alpha} e^{i\psi} \right] \sin(2 x) = |m_{\alpha\alpha}| \cos (2 (x - x_\alpha)),
\end{align}
Here $x_\alpha$ and $y_\alpha$ don't depend on $x, y$. They are introduced for convenience and can be defined by equations above.

Inequations \eqref{eq:notional_close} follow directly from expressions above.

We start with the case:
\begin{equation}
\label{eq:limit_app}
|\mathcal{M}_{2 \alpha} -\mathcal{M}_{1 \alpha}| \ll \mathcal{M}_{1 \alpha} + \mathcal{M}_{2 \alpha}
\end{equation}
In this limit $\mathcal{M}_{1 \alpha} \approx \mathcal{M}_{2 \alpha}$.
As such, equation \eqref{eq:2HNL_y} can be solved for variable $y$ (as a function of, for example, $\mathcal{M}_{1 e}$):
\begin{align}
\label{eq:sinh}
\sinh(2 y) =& - \frac{2 \mathcal{M}_{1 e}\times 2 \Im \left[ \lambda_{11}^* \lambda_{21} e^{i\psi} \right] }{|m_{ee}|^2} \pm \frac{(|\lambda_{11}|^2 + |\lambda_{21}|^2) \sqrt{4 \mathcal{M}_{1 e}^2 - |m_{ee}|^2}}{|m_{ee}|^2}\\
\label{eq:cosh}
\cosh(2 y) =&   \frac{2 \mathcal{M}_{1 e} (|\lambda_{11}|^2 + |\lambda_{21}|^2)}{|m_{ee}|^2}\pm  \frac{2 \Im \left[ \lambda_{11}^* \lambda_{21} e^{i\psi} \right] \sqrt{4 \mathcal{M}_{1 e}^2 - |m_{ee}|^2}}{|m_{ee}|^2} 
\end{align}

But we can consider a bit stricter limit than \eqref{eq:limit_app}:
\begin{equation}
\label{eq:limit2_app}
\mathcal{M}_{1 \alpha} + \mathcal{M}_{2 \alpha} \gg |m_{\alpha\alpha}| \geq |\mathcal{M}_{2 \alpha} - \mathcal{M}_{1 \alpha}|
\end{equation}

In this limit \eqref{eq:sinh} and \eqref{eq:cosh} take even simpler form:
\begin{align}
\label{eq:sinh2}
\sinh(2 y) \approx& - 2 \mathcal{M}_{1 e} \frac{ 2 \Im \left[ \lambda_{11}^* \lambda_{21} e^{i\psi} \right] \pm (|\lambda_{11}|^2 + |\lambda_{21}|^2)}{|m_{ee}|^2}\approx \mp \cosh(2y)
\end{align}

As a result, one can write:
\begin{align}
\label{eq:Ualpha}
\frac{|U_{i \alpha}|^2}{|U_{i\,tot}|^2} =& \frac{\left(|\lambda_{1\alpha}|^2 + |\lambda_{2\alpha}|^2 \mp 2 \Im \left[ \lambda_{1\alpha}^* \lambda_{2\alpha} e^{i\psi} \right]\right)}{\sum_{j=1}^3 \left(|\lambda_{1j}|^2 + |\lambda_{2j}|^2 \mp 2 \Im \left[\lambda_{1j}^* \lambda_{2j} e^{i\psi} \right]\right)}
\end{align}
where $|U_{i\,tot}|^2 = |U_{i e}|^2 + |U_{i \mu}|^2 + |U_{i \tau}|^2$.

In limit \eqref{eq:limit_app} expression \eqref{eq:Ualpha} take somewhat more cumbersome form:
\begin{align}
\label{eq:Ualpha_w_kappa}
\frac{|U_{i \alpha}|^2}{|U_{i\,tot}|^2} =& \Big(\left(|\lambda_{11}|^2 + |\lambda_{21}|^2\right)\left(|\lambda_{1\alpha}|^2 + |\lambda_{2\alpha}|^2 \right) -4\Im \left[ \lambda_{11}^* \lambda_{21} e^{i\psi} \right] \Im \left[ \lambda_{1\alpha}^* \lambda_{2\alpha} e^{i\psi} \right] \pm \nonumber\\
& 2 \kappa_i \Big(\Im \left[ \lambda_{11}^* \lambda_{21} e^{i\psi} \right] \left(|\lambda_{1\alpha}|^2 + |\lambda_{2\alpha}|^2 \right) -\left(|\lambda_{11}|^2 + |\lambda_{21}|^2\right) \Im \left[ \lambda_{1\alpha}^* \lambda_{2\alpha}  e^{i\psi} \right]\Big) \Big)\times\nonumber\\
&\Big(\left(|\lambda_{11}|^2 + |\lambda_{21}|^2\right)\sum_{j=1}^3\left(|\lambda_{1j}|^2 + |\lambda_{2j}|^2\right) -4\Im \left[\lambda_{11}^* \lambda_{21} e^{i\psi} \right] \Im \left[\sum_{j=1}^3\left(\lambda_{1j}^* \lambda_{2j}\right) e^{i\psi} \right] \pm \nonumber\\
&  2 \kappa_i \Big(\Im \left[ \lambda_{11}^* \lambda_{21} e^{i\psi} \right] \sum_{j=1}^3\left(|\lambda_{1j}|^2 + |\lambda_{2j}|^2\right) - \left(|\lambda_{11}|^2 + |\lambda_{21}|^2\right) \Im \left[\sum_{j=1}^3\left(\lambda_{1j}^* \lambda_{2j}\right) e^{i\psi} \right]\Big) \Big)^{-1}\nonumber
\end{align}
where $\kappa_i = \sqrt{1 - \frac{|m_{ee}|^2}{4\mathcal{M}_{i e}^2}}, 0< \kappa < 1$.
Limit \eqref{eq:limit2_app} corresponds to $\kappa \to 1$, defining values $\mathcal{M}_{i e}$ much greater than their minimal value.

\section{Three HNLs case}
\label{app:3HNL}
In this section, we show that similar results can be achieved for the three HNLs case.

We use the following parametrization of $R$:
\begin{eqnarray}
\label{eq5}
&&R= diag\{\pm 1, \pm 1, \pm 1\} \left( \begin{array}{ccc}
c_2 c_1 & c_2 s_1 & s_2\\
-c_3 s_1 - s_3 s_2 c_1 & c_3 c_1 - s_3 s_2 s_1 & s_3 c_2\\
s_3 s_1 - c_3 s_2 c_1 & -s_3 c_1 - c_3 s_2 s_1 & c_3 c_2\\
\end{array}\right)
\end{eqnarray}
Here $c_i =\cos z_i, s_i =\sin z_i, i=\overline{1,3}$, and $z_i \subset \mathbb{C}$ are not restricted in any way.

To further simplify our calculations, we are going to use the following variables much the same way we did it for the two HNLs case:
\begin{align}
A \equiv  & diag\{e^{i \frac{\alpha_1}{2}}; e^{i \frac{\alpha_2}{2}}; 1\} \times U_{PMNS}^\dagger\\
\lambda_{1i}  \equiv & \sqrt{m_1} A_{1i} c_1 + \sqrt{m_2} A_{2i} s_1\\
\lambda_{2i}  \equiv & \sqrt{m_3} A_{3i}\\
\lambda_{3i}  \equiv & - \sqrt{m_1} A_{1i} s_1 + \sqrt{m_2} A_{2i} c_1\\
\Gamma_{1i}  \equiv & \lambda_{1i} c_2 + \lambda_{2i} s_2\\
\Gamma_{4i}  \equiv & \lambda_{2i} c_2 - \lambda_{1i} s_2\\
\Gamma_{2i}  \equiv & \lambda_{3i} c_3 + \Gamma_4 s_3\\
\Gamma_{3i}  \equiv & \Gamma_4 c_3 - \lambda_{3i} s_3
\end{align}
One can note that $\lambda_{1i}^2 + \lambda_{3i}^2 = m_1 A_{1i}^2 + m_2 A_{2i}^2$ does not depend on $z_1$.
In the same way $\Gamma_{1i}^2 + \Gamma_{4i}^2 = \lambda_{1i}^2 + \lambda_{2i}^2$ does not depend on $z_2$ and $\Gamma_{2i}^2 + \Gamma_{3i}^2 = \Gamma_{4i}^2 + \lambda_{3i}^2$ does not depend on $z_3$.

The mixing matrix can be expressed as:
\begin{eqnarray}
\label{eq7}
&&U =  i \times diag\{\pm 1, \pm 1, \pm 1\} \left( \begin{array}{ccc}
\frac{1}{\sqrt{M_1}} \Gamma_{11} & \frac{1}{\sqrt{M_1}} \Gamma_{12} & \frac{1}{\sqrt{M_1}} \Gamma_{13}\\
\frac{1}{\sqrt{M_2}} \Gamma_{21} & \frac{1}{\sqrt{M_2}} \Gamma_{22} & \frac{1}{\sqrt{M_2}} \Gamma_{23}\\
\frac{1}{\sqrt{M_3}} \Gamma_{31} & \frac{1}{\sqrt{M_3}} \Gamma_{32} & \frac{1}{\sqrt{M_3}} \Gamma_{33}\\
\end{array}\right)
\end{eqnarray}
Note that, once again, we are interested only in the squared elements of mixing matrix $U$, therefore both the $i$ and the signs matrix don't affect expressions $|U_{ij}|^2$.

\subsection{Lower limit for mixing with specific flavor}
\label{app:minimal U_e}

First of all, we study the sum of squared mixing of all three HNLs with a specified active neutrino flavour.
The same reasoning applies here as the one we used in two HNLs case: the sum is a viable parameter for experimental search by itself and at least one HNL mixing angle is guaranteed to be no less then one third of the sum.
%in three HNL case we take the arithmetic mean of mixing of all three HNL with one flavor as the minimal value.
%The reasoning is the same as when we have taken half sum in two HNL case: we know with certainty that mixing of at least one HNL with that flavor is equal to or greater than that minimal value.
\begin{align}
\label{eq:U_alpha}
U_{sum, \alpha^2} \equiv& |U_{1\alpha}|^2 + |U_{2\alpha}|^2 + |U_{3\alpha}|^2 = \frac{1}{M_1} |\Gamma_{1\alpha}|^2 + \frac{1}{M_2} |\Gamma_{2\alpha}|^2 + \frac{1}{M_3} |\Gamma_{3\alpha}|^2
\end{align}

The $z_3$ extremum criterion is:
\begin{equation}
\label{eq:dz3}
\frac{\partial U^2_{sum, \alpha}}{\partial z_3}  =  \frac{1}{M_2} \Gamma_{2\alpha}^*\Gamma_{3\alpha} -  \frac{1}{M_3} \Gamma_{3\alpha}^*\Gamma_{2\alpha} = 0
\end{equation}

This equation can have three solutions:
\begin{align}
\label{eq:Gamma_2=0}
\Gamma_{2\alpha}  = & 0\\
\label{eq:Gamma_3=0}
\Gamma_{3\alpha}  = & 0\\
\label{eq:M_2=M_3}
M_2  = & M_3, \Gamma_{2\alpha}^*\Gamma_{3\alpha} - \Gamma_{2\alpha}\Gamma_{3\alpha}^* = 0 
\end{align}
In last equation we consider $ |\Gamma_{2\alpha}| \neq 0, |\Gamma_{3\alpha}| \neq 0$.

When $\Gamma_{2\alpha}=0$ it automatically means that $\Gamma_{3\alpha}^2 = \Gamma_{2\alpha}^2 + \Gamma_{3\alpha}^2 = \Gamma_{4\alpha}^2 + \lambda_{3\alpha}^2$.
When $\Gamma_{3\alpha}=0$ it automatically means that $\Gamma_{2\alpha}^2 = \Gamma_{2\alpha}^2 + \Gamma_{3\alpha}^2 = \Gamma_{4\alpha}^2 + \lambda_{3\alpha}^2$.
Therefore, in accordance with the definition \eqref{eq:U_alpha}:
\begin{equation}
\label{eq:U_e_z_3_extremum}
U^2_{sum, \alpha|_{\frac{\partial U^2_\alpha}{\partial z_3}=0}, M_2\neq M_3} =  \frac{1}{M_1} |\Gamma_{1\alpha}|^2 + \frac{1}{M} |\Gamma_{4\alpha}^2 + \lambda_{3\alpha}^2|,
\end{equation}
where $M=max\{M_2,M_3\}$.
In other words, solutions for \eqref{eq:Gamma_2=0} and \eqref{eq:Gamma_3=0} are identical except for the factors $\frac{1}{M_2}$ and $\frac{1}{M_3}$ before the value $|\Gamma_{4\alpha}^2 + \lambda_{3\alpha}^2|$.

If $M_2 = M_3 \equiv M$ the eq. \eqref{eq:U_alpha} takes form:
\begin{equation}
\label{eq:(M_2=M_3)U_e}
U_{sum, \alpha}^2 =  \frac{1}{M_1} |\Gamma_{1\alpha}|^2 + \frac{1}{M} \left(|\Gamma_{2\alpha}|^2 + |\Gamma_{3\alpha}|^2\right)
\end{equation}
Simplifying expression $|\Gamma_{2\alpha}|^2 + |\Gamma_{3\alpha}|^2$ one can obtain:

\begin{align}
\label{eq:Gamma_2+Gamma_3}
|\Gamma_{2\alpha}|^2 + |\Gamma_{3\alpha}|^2 =& \left(|\lambda_{3\alpha}|^2+|\Gamma_{4\alpha}|^2\right) \cosh(2 y_3) + 2\,\textrm{Im}[\lambda_{3\alpha}\Gamma_{4\alpha}^*] \sinh(2 y_3)
\end{align}

Therefore, if $M_2 = M_3$, $U_\alpha^2$ doesn't depend on $x_3$ anymore.
The $y_3$ extremum condition is:
\begin{align}
\label{eq:y3_extremum}
\left(|\lambda_{3\alpha}|^2+|\Gamma_{4\alpha}|^2\right) \sinh(2 y_3) +2\,\Im[\lambda_{3\alpha}\Gamma_{4\alpha}^*] \cosh(2 y_3) = 0
\end{align}

After simplifying the eq. \eqref{eq:M_2=M_3}, one can notice that it can be reduced to the equation \eqref{eq:y3_extremum}.

Taking into account \eqref{eq:Gamma_2+Gamma_3} and \eqref{eq:y3_extremum}, and solving them with respect to $y_3$, one obtains:
\begin{align}
\label{eq:(G2+G3)_y_3_extremum}
\left(|\Gamma_{2\alpha}|^2 + |\Gamma_{3\alpha}|^2\right)|_{\frac{\partial U^2_\alpha}{\partial z_3}=0} &=\sqrt{\left(|\lambda_{3\alpha}|^2+|\Gamma_{4\alpha}|^2\right)^2 - 4\left(\Im[\lambda_{3\alpha}\Gamma_{4\alpha}^*]\right)^2} \nonumber\\
&\equiv |\Gamma_{4\alpha}^2 + \lambda_{3\alpha}^2|
\end{align}
Combining \eqref{eq:(M_2=M_3)U_e} and \eqref{eq:(G2+G3)_y_3_extremum} one obtains the same result as in \eqref{eq:U_e_z_3_extremum}.

The next step is to find the $z_2$ extremum of eq. \eqref{eq:U_e_z_3_extremum}:
\begin{align}
\label{eq:dz2 explained}
&\frac{1}{M_1} \Gamma_{1\alpha}^*\Gamma_{4\alpha} - \frac{1}{M} \Gamma_{1\alpha} \Gamma_{4\alpha} e^{- 2 i \gamma_{2\alpha}} = 0,\\
&\textrm{ where } e^{ 2 i \gamma_{2\alpha}}  = \frac{\lambda_{3\alpha}^2 + \Gamma_{4\alpha}^2}{|\lambda_{3\alpha}^2 + \Gamma_{4\alpha}^2|} \nonumber
\end{align}

Similarly to eq. \eqref{eq:dz3} it has three solutions:
\begin{align}
\label{eq:Gamma_1=0}
\Gamma_{1\alpha} = & 0\\
\label{eq:Gamma_4=0}
\Gamma_{4\alpha} = & 0\\
\label{eq:M_1=M_3}
M_1 = & M, \Gamma_{1\alpha} \equiv |\Gamma_{1\alpha}| e^{i\gamma_{1\alpha}}, e^{2 i\left(\gamma_{1\alpha}-\gamma_{2\alpha}\right)} = 1 
\end{align}
In the last equation, once again, $|\Gamma_{1\alpha}|~\neq~0, |\Gamma_{4\alpha}|~\neq~0$.

Taking into account that $\Gamma_{1\alpha}^2 + \Gamma_{4\alpha}^2 = \lambda_{1\alpha}^2 + \lambda_{2\alpha}^2$ we obtain from eq. \eqref{eq:U_e_z_3_extremum}:
\begin{align}
\label{eq:Gamma_1=0 explained}
&\Gamma_{1\alpha}  =  0   \Rightarrow U^2_{sum, \alpha|_{\frac{\partial U^2_\alpha}{\partial z_3}=0}, \Gamma_{1\alpha}  =  0} =  \frac{1}{M} |\lambda_{1\alpha}^2 + \lambda_{2\alpha}^2 + \lambda_{3\alpha}^2| \\
&\Gamma_{4\alpha} = 0 \Rightarrow U^2_{sum, \alpha|_{\frac{\partial U^2_\alpha}{\partial z_3}=0}, \Gamma_{4\alpha}  =  0} = \frac{1}{M_1}|\lambda_{1\alpha}^2 + \lambda_{2\alpha}^2 | + \frac{1}{M} |\lambda_{3\alpha}|^2 \label{eq:Gamma_4=0 explained}
\end{align}

Eq. \eqref{eq:Gamma_1=0 explained} doesn't depend on $z_1$.
Minimizing eq. \eqref{eq:Gamma_4=0 explained} over $z_1$ we obtain solutions $\lambda_{1\alpha} = 0$ and $\lambda_{3\alpha}=0$ and a special case for $M_1=M$.
Obviously, if $M_1=M_{max}\geq M$, then for the minimal $U_\alpha^2$ we take $\lambda_{3\alpha}=0$ and obtain $U_{\alpha|_{\frac{\partial U^2_\alpha}{\partial z_3}=0}, \Gamma_{1\alpha}  =  0, \lambda_{3\alpha}  =  0} =  \frac{1}{M_{max}} |\lambda_{1\alpha}^2 + \lambda_{2\alpha}^2|$.
If $M_1<M=M_{max}$, then both solutions $(\frac{1}{M_1}| \lambda_{2\alpha} |^2 + \frac{1}{M} |\lambda_{3\alpha}|^2)_{|_{\lambda_{1\alpha}=0}}$ and $(\frac{1}{M_1}|\lambda_{1\alpha}^2 + \lambda_{2\alpha}^2 |)_{|_{\lambda_{3\alpha}=0}}$ yield us expressions that are greater than or equal to the solution \eqref{eq:Gamma_1=0 explained}.
Taking into account that $|\lambda_{1\alpha}^2 + \lambda_{2\alpha}^2 + \lambda_{3\alpha}^2 | = |m_{\alpha\alpha}|$, we obtain that $U_{min, \alpha, M_1\neq M}^2 = \frac{|m_{\alpha\alpha}|}{M_{max}}$.

Let's check the final case $M_1=M$. The criterion for the minimum on $z_1$ for the expression  \eqref{eq:U_e_z_3_extremum}:
\begin{equation}
\frac{1}{M} \lambda_{3\alpha} \left(c_2 \Gamma_{1\alpha}^* - \left(s_2 \Gamma_{4\alpha} + \lambda_{1\alpha}\right)e^{-2 i \gamma_{2\alpha}}\right)= 0
\end{equation}
One can notice that $\Gamma_{4\alpha} s_2  + \lambda_{1\alpha} = \lambda_{2\alpha} s_2 c_2 - \lambda_{1\alpha} s_2^2 + \lambda_{1\alpha} = c_2 (\lambda_{1\alpha} c_2 + \lambda_{2\alpha} s_2) = c_2 \Gamma_{1\alpha}$.
Therefore, remembering from eq. \eqref{eq:M_1=M_3} that $e^{2 i\gamma_{2\alpha}} = e^{2 i\gamma_{1\alpha}}$ we obtain:
\begin{equation}
\frac{1}{M} \lambda_{3\alpha} c_2 \left(\Gamma_{1\alpha}^* -\Gamma_{1\alpha} e^{-2 i\gamma_{1\alpha}})\right) \equiv 0\nonumber
\end{equation}
This expression is equivalent to zero and therefore the sole restriction is the eq. \eqref{eq:M_1=M_3}:
\begin{equation}
\Gamma_{1\alpha}^* -\Gamma_{1\alpha} e^{-2 i\gamma_{2\alpha}} = 0
\end{equation}
The expression for the minimal $U^2_{min, \alpha}$, in this case, can be rewritten as:
\begin{align}
U^2_{min, \alpha} =& \frac{1}{M} \left(|\Gamma_{1\alpha}^2| + |\Gamma_{4\alpha}^2 + \lambda_{3\alpha}^2| \right)=  \frac{1}{M} \left(\Gamma_{1\alpha}^2 + \Gamma_{4\alpha}^2 + \lambda_{3\alpha}^2 \right)
 e^{-2 i\gamma_{2\alpha}} \nonumber\\
= & \frac{1}{M} \left(\lambda_{1\alpha}^2 + \lambda_{2\alpha}^2 + \lambda_{3\alpha}^2 \right)
 e^{-2 i\gamma_{2\alpha}}
\end{align}

The expression remains real, and we can take the absolute value of both the left and the right parts of the equation above and once again obtain:
\begin{align}
\label{eq:U_e_min_app}
U^2_{min, \alpha} =&  \frac{|m_1 A_{1\alpha}^2 + m_2 A_{2\alpha}^2 + m_3 A_{3\alpha}^2|}{M}  \equiv \frac{|m_{\alpha\alpha}|}{M}
\end{align}
Here $M=max\{M_1, M_2, M_3\}$. 
Absolutely the same approach is used to investigate the $M_1=M$ case in \eqref{eq:Gamma_4=0 explained}, giving us the same answer.

Note that to obtain eq. \eqref{eq:U_e_min_app} for a specified $\alpha$ we fixate the values of $z_1, z_2, z_3$.
Therefore, choosing $U^2_e=U^2_{e_{min}}$ we also affect the values of $U^2_\mu$, $U^2_\tau$.
One has to check that the values of $U^2_\mu$ and $U^2_\tau$ obtained in this way lay in the experimentally unrestricted area. 

\subsection{Pseudodegenerate state for three HNLs}
\label{app:3HNL_pseudo}

We can write down similar relations between notional masses $\mathcal{M}_{i\alpha} = M_i |U_{i\alpha}|^2$ and $|\Gamma_{1\alpha}|^2 , |\Gamma_{2\alpha}|^2 , |\Gamma_{3\alpha}|^2 $ to the ones we had in two-HNLs case:
\begin{align}
\label{eq:3HNL_y3}
\mathcal{M}_{3\alpha} +  \mathcal{M}_{2\alpha} =& |\Gamma_{1\alpha}^2 - m_{\alpha\alpha}| \cosh (2 (y_3 - \hat{y}_\alpha)),\\
\label{eq:3HNL_x3}
\mathcal{M}_{3\alpha} -  \mathcal{M}_{2\alpha} =& |\Gamma_{1\alpha}^2 - m_{\alpha\alpha}|  \cos (2 (x_3 - \hat{x}_\alpha)),\\
|\mathcal{M}_{1\alpha} -  |m_{\alpha\alpha}|| \leq& |\Gamma_{1\alpha}^2 - m_{\alpha\alpha}| \,\leq\,  \mathcal{M}_{1\alpha} +  |m_{\alpha\alpha}|,
\end{align}
here $\hat{x}_\alpha, \hat{y}_\alpha$ don't depend on $x_3, y_3$.

Knowing that $\cos(X)\leq 1, \cosh(Y) \geq 1$ for real $X,Y$, we can confer that mixing satisfies the following criteria:
\begin{align}
|\mathcal{M}_{3 \alpha} - &\mathcal{M}_{2 \alpha}| - |m_{\alpha\alpha}| \leq   \mathcal{M}_{1 \alpha} \, \leq \, \mathcal{M}_{2 \alpha} + \mathcal{M}_{3 \alpha} + |m_{\alpha\alpha}|\\
|\mathcal{M}_{3 \alpha} - &\mathcal{M}_{1 \alpha}| - |m_{\alpha\alpha}| \leq \mathcal{M}_{2 \alpha} \, \leq \, \mathcal{M}_{1 \alpha} + \mathcal{M}_{3 \alpha} + |m_{\alpha\alpha}|\\
|\mathcal{M}_{2 \alpha} - &\mathcal{M}_{1 \alpha}| - |m_{\alpha\alpha}| \leq \mathcal{M}_{3 \alpha} \, \leq \, \mathcal{M}_{1 \alpha} + \mathcal{M}_{2 \alpha} + |m_{\alpha\alpha}|
\end{align}

We are interested in the following limit:
\begin{equation}
\label{eq:3HNL_requirements_app}
|\mathcal{M}_{3 \alpha}- \mathcal{M}_{2 \alpha}| \ll \mathcal{M}_{2 \alpha} + \mathcal{M}_{3 \alpha},
\end{equation}
The underlining idea is the same: pseudodegenerate state is achieved when in eqs. \eqref{eq:3HNL_y3}, \eqref{eq:3HNL_x3} we have $\cosh (2 (y_3 - \hat{y_\alpha})) \gg \cos (2 (x_3 - \hat{x_\alpha}))$.
It automatically holds true for $\cosh (2 (y_3 - \hat{y_\alpha})) \gg 1$ or $\cos (2 (x_3 - \hat{x_\alpha})) \ll 1$.
The two cases give us inequations \eqref{eq:3HNL_requirements} and \eqref{eq:same_magnitude_massmixing} we mention in the main text.
For three HNLs case the ``scale factor'' in eqs. \eqref{eq:3HNL_y3}, \eqref{eq:3HNL_x3} depends not only on $|m_{\alpha\alpha}|$, but also on $\mathcal{M}_{1 \alpha}$.

To achieve the pseudodegenerate state, we need to obtain the expressions for the ratio $|U_e|^2:|U_\mu|^2:|U_\tau|^2$. It now depends not only on masses and yet-unknown CP-violating phases $\delta, \alpha_1\alpha_2$, but also on HNL sector anles $z_1, z_2$:
\begin{align}
\label{eq:Ue_3HNL}
\frac{|U_{i \alpha}|^2}{|U_{i\,tot}|^2} =& \frac{\left(|\lambda_{3\alpha}|^2 + |\Gamma_{4\alpha}|^2 \mp 2 \Im \left[ \lambda_{3\alpha}^* \Gamma_{4\alpha} \right]\right)}{\sum_{j=1}^3\left(|\lambda_{3j}|^2 + |\Gamma_{4j}|^2 \mp 2 \Im \left[\lambda_{3j}^* \Gamma_{4j} \right]\right)}
\end{align}

Values of $|\Gamma_{1\alpha}|^2$ are defined by the same parameters that we scan in eq. \eqref{eq:Ue_3HNL}.
Therefore, if we fixate this ratio and the mass $M_1$, we obtain $|U_{1\alpha}|^2 = \frac{|\Gamma_{1\alpha}|^2}{M_1}$.
We still have freedom of choice for $z_3$ parameters, so inequation \eqref{eq:3HNL_requirements_app} doesn't restrict possible values for the masses $M_2, M_3$ in pseudodegenerate state.
At the same time $|U_{2\alpha}|^2 \gg \frac{M_1}{M_2} |U_{1\alpha}|^2$ and $|U_{3\alpha}|^2 \gg \frac{M_1}{M_3} |U_{1\alpha}|^2$ should still lay in areas that haven't been restricted by existing experimental searches.
That automatically holds true in the case of hierarchy $M_1 \ll M_2, M_3$, $|U_{1\alpha}|^2 \sim |U_{2\alpha}|^2 \sim |U_{3\alpha}|^2$, traditionally used for models where lightest HNL serves the role of a dark matter particle, including $\nu MSM$~\cite{Akhmedov:1998qx,Asaka:2011pb}.

\end{appendices}


\end{document}